\documentclass[apj,revtex4]{emulateapj}
\usepackage{multirow}
\usepackage{graphics}
\usepackage{natbib}
\usepackage{footnote}
\usepackage{amsmath,amsthm}
\usepackage{longtable}
\usepackage[breaklinks,colorlinks,urlcolor=blue,citecolor=blue,linkcolor=blue]{hyperref} 
\usepackage{ulem}
\usepackage{gensymb}
\usepackage{amsmath}	
\usepackage{color}

\shorttitle{The 105 Month \textit{Swift}-BAT All-Sky Hard X-ray Survey}
\shortauthors{Oh et al.}	

\begin{document}

\title{The 105 Month \textit{Swift}-BAT All-Sky Hard X-ray Survey}

\author{Kyuseok Oh\altaffilmark{1,2,9}, Michael Koss\altaffilmark{2,3,10}, Craig B. Markwardt\altaffilmark{4}, Kevin Schawinski\altaffilmark{2}, Wayne H. Baumgartner\altaffilmark{4}, Scott D. Barthelmy\altaffilmark{4}, S. Bradley Cenko\altaffilmark{4}, Neil Gehrels\altaffilmark{4},  Richard Mushotzky\altaffilmark{5}, Abigail Petulante\altaffilmark{5}, Claudio Ricci\altaffilmark{6}, Amy Lien\altaffilmark{7,8}, Benny Trakhtenbrot\altaffilmark{2,11}}

\altaffiltext{1}{Department of Astronomy, Kyoto University, Oiwake-cho, Sakyo-ku, Kyoto 606-8502, Japan; ohk@kusastro.kyoto-u.ac.jp}
\altaffiltext{2}{Institute for Particle Physics and Astrophysics, Department of Physics, ETH Zurich, Wolfgang-Pauli-Strasse 27, CH-8093 Zurich, Switzerland}
\altaffiltext{3}{Eureka Scientific Inc., 2452 Delmer Street, Suite 100, Oakland, CA 94602-3017, USA}
\altaffiltext{4}{NASA Goddard Space Flight Center, Greenbelt, MD 20771, USA}
\altaffiltext{5}{Astronomy Department, University of Maryland, College Park, MD, 20742, USA}
\altaffiltext{6}{Instituto de Astrof\'{\i}sica, Facultad de F\'{\i}sica, Pontificia Universidad Cat\'olica de Chile, Casilla 306, Santiago 22, Chile}
\altaffiltext{7}{Center for Research and Exploration in Space Science and Technology (CRESST) and NASA Goddard Space Flight Center, Greenbelt, MD 20771, USA}
\altaffiltext{8}{Department of Physics, University of Maryland, Baltimore County, 1000 Hilltop Circle, Baltimore, MD 21250, USA}
\altaffiltext{9}{\textrm{JSPS fellow}}
\altaffiltext{10}{\textrm{Ambizione fellow}}
\altaffiltext{11}{\textrm{Zwicky fellow}}

\def\NH{$N_{\textrm{H}}$}
\def\Ebmv{E($B-V$)}
\def\LOIII{$L[\mbox{O\,{\sc iii}}]$}
\def\Ledd{${L/L_{\rm Edd}}$}
\def\LOIIIs4{$L[\mbox{O\,{\sc iii}}]$/$\sigma^4$}
\def\LOIIIMbh{$L[\mbox{O\,{\sc iii}}]$/$M_{\rm BH}$}
\def\Mbh{$M_{\rm BH}$}
\def\Msigma{$M_{\rm BH} - \sigma$}
\def\Ms{$M_{\rm *}$}
\def\Msun{$M_{\odot}$}
\def\Msunyr{${\rm M_{\odot}yr^{-1}}$}

\def\kms{${\rm km}~{\rm s}^{-1}$}
\newcommand{\cms}{\mbox{${\rm cm\;s^{-1}}$}}
\newcommand{\pccm}{\mbox{${\rm cm^{-3}}$}}
\newcommand{\ergs}	{\ifmmode {\rm erg\,s}^{-1} \else erg s$^{-1}$\fi}
\newcommand{\ergcms}	{\ifmmode {\rm erg\,cm}^{-2}\,{\rm s}^{-1} \else erg\,cm$^{-2}$\,s$^{-1}$\fi}

\newcommand{\senshalf}{$7.24\times 10^{-12}\ {\rm erg\ s^{-1}\ cm^{-2}}$}			
\newcommand{\sensninety}{$8.40\times 10^{-12}\ {\rm erg\ s^{-1}\ cm^{-2}}$}			

\newcommand{\Nnewtotal}{422} 
\newcommand{\NnewBeamed}{43} 
\newcommand{\fracnewBeamed}{10} 
\newcommand{\NnewSy}{144} 
\newcommand{\fracnewSy}{34} 
\newcommand{\NnewXRB}{31} 
\newcommand{\fracnewXRB}{7} 
\newcommand{\NnewIdentified}{328} 
\newcommand{\Nunidentified}{94} 

\newcommand{\Ntotal}{1632} 
\newcommand{\NtotAGN}{947} 
\newcommand{\Nunknown}{129} 
\newcommand{\Nmultiple}{10} 

\newcommand{\NUnone}{36} 
\newcommand{\NUntwo}{55}
\newcommand{\NUnthree}{38}  
\newcommand{\NUnknown}{129}

\newcommand{\NSyone}{379}
\newcommand{\NSytwo}{448} 
\newcommand{\NLINER}{6}     
\newcommand{\Nbeamed}{158}
\newcommand{\NCV}{75}
\newcommand{\NSymb}{4}
\newcommand{\Notherstar}{12}   
\newcommand{\NPulsar}{25}
\newcommand{\NSNR}{7}
\newcommand{\NNova}{6}
\newcommand{\NHMXB}{108}
\newcommand{\NLMXB}{109}
\newcommand{\NXRB}{8}
\newcommand{\Ncluster}{26}
\newcommand{\NUnknownAGN}{114}
\newcommand{\NUnknownall}{139} 

\newcommand{\NBATmatchedwithPalermo}{1125}	
\newcommand{\NBATnewmatchedwithPalermo}{102} 
\newcommand{\fracbatmatchedwithPalermo}{68} 
\newcommand{\NAGNmatchedwithPalermo}{777} 
\newcommand{\NXRBmatchedwithPalermo}{192} 
\newcommand{\NAGNXRBmatchedwithPalermo}{969} 
\newcommand{\NexclusivelyinBAT}{507} 
\newcommand{\NAGNexclusivelyinBAT}{347}  
\newcommand{\NunknownexclusivelyinBAT}{109} 
\newcommand{\NexclusivelyinPalermo}{159} 
\newcommand{\fracsofterband}{65} 
\newcommand{\Nsofterband}{104} 

\definecolor{myblue}{RGB}{0, 100, 220}

\begin{abstract}
We present a catalog of hard X-ray sources detected in the first 105 months of observations with the Burst Alert Telescope (BAT) coded mask imager on board the \textit{Swift} observatory. The 105 month \textit{Swift}-BAT survey is a uniform hard X-ray all-sky survey with a sensitivity of \sensninety\ over 90\% of the sky and \senshalf\ over 50\% of the sky in the $14-195$ keV band. The \textit{Swift}-BAT 105 month catalog provides \Ntotal\ (\Nnewtotal\ new detections) hard X-ray sources in the $14-195$ keV band above the $4.8\sigma$ significance level. Adding to the previously known hard X-ray sources, \fracnewSy \% (\NnewSy/\Nnewtotal) of the new detections are identified as Seyfert AGN in nearby galaxies ($z<0.2$). The majority of the remaining identified sources are X-ray binaries (\fracnewXRB \%, \NnewXRB) and blazars/BL Lac objects (\fracnewBeamed \%, \NnewBeamed). As part of this new edition of the \textit{Swift}-BAT catalog, we release eight-channel spectra and monthly sampled light curves for each object in the online journal and at the \textit{Swift}-BAT 105 month Web site. 
\end{abstract}	


\section{Introduction}
Since the first X-ray satellite (\textit{Uhuru}, \citealt{Giacconi71}) launched on 1970, a large number of surveys has been made in both the soft and hard X-ray bands with extensive follow-up analysis.  \citet{Forman78} presented 339 X-ray sources observed with the \textit{Uhuru} X-ray observatory in the $2-20$ keV energy band. Later, \textit{HEAO}-A4, the X-ray and gamma-ray instrument onboard the \textit{HEAO 1} satellite, conducted all-sky survey in the $13-180$ keV range from 1977 to 1979 detecting 77 sources in its catalog \citep{Levine84}. 

Compared to the soft X-ray energy band, a hard X-ray all-sky survey ($>10$ keV) provides an important way of studying astrophysical objects since such energetic hard X-ray photons can pass through large columns of gas and dust detecting even Compton-thick sources ($N_{\rm H}>10^{24} {\rm cm^{-2}}$, \citealt{Ricci15, Koss16}). 

The \textit{Swift} gamma-ray burst (GRB) observatory \citep{Gehrels04}, which was launched in 2004 November, and is successfully carrying on a all-sky hard X-ray survey at $14-195$ keV with the Burst Alert Telescope (BAT). The \textit{Swift} GRB observatory is primarily designed to detect transient GRBs with a coded-mask telescope \citep{Barthelmy05} which has a very wide field of view ($\sim60\degree\times100\degree$). When the BAT discovers a new GRB, two narrow field instruments, the X-ray Telescope (XRT; \citealt{Burrows05}) and the Ultraviolet/optical Telescope (UVOT; \citealt{Roming05}), observe the GRB candidate \citep{Sakamoto08, Sakamoto11, Lien16}. Based on the stacking of these and other observations, \textit{Swift} has been successfully carrying out an all-sky survey. 

\begin{figure*}[t!]
\centering
	\includegraphics[width=\linewidth]{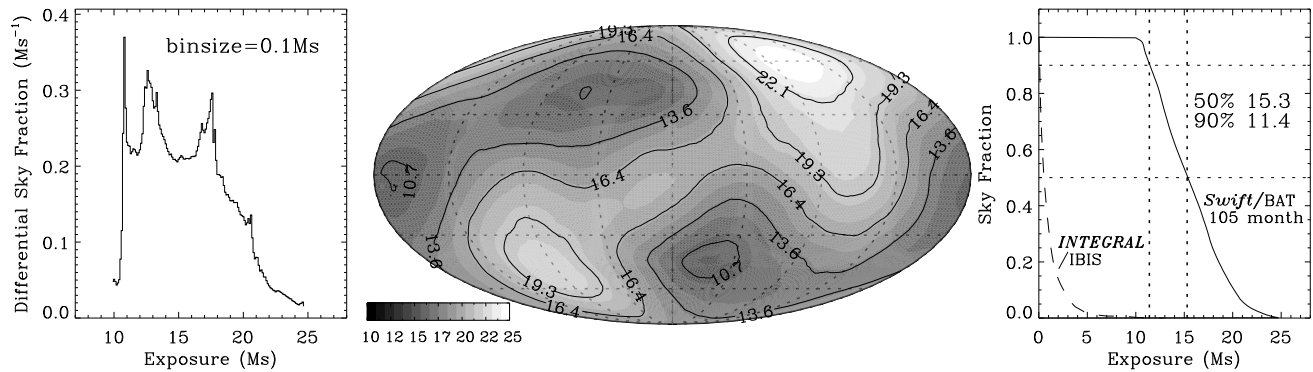}
    \caption{105 month $Swift$-BAT survey all-sky exposure. The left panel shows the distribution of exposure times across the sky, the middle panel shows an all-sky exposure map in a galactic projection, and the right panel shows the fraction of sky covered as a function of exposure time. A half of the sky (50\%) is observed with more than 15.3 Ms of exposure time, while 90\% of the sky is covered with 11.4 Ms. Fraction of all sky covered by \textit{INTEGRAL}-IBIS \citep{Bird16} as a function of exposure time is shown with dashed line as a comparison.}
    \label{fig1}
\end{figure*}

The first \textit{Swift}-BAT survey catalog used the initial 3 months of observations to study high Galactic latitude sources \citep{Markwardt05}. \citet{Tueller08} focused on active galactic nuclei (AGN) observed from the first 9 months of data, and later \citet{Tueller10} provided a catalog of the first 22 months of data. \citet{Baumgartner13} published the eight-channel spectra as well as monthly light curves from the first 70 months of data with improved data processing.  Independent efforts have also been made on the analysis of 39 month \citep{Cusumano10a}, 54 month \citep{Cusumano10b}, 66 month, and 100 month \textit{Swift}-BAT data by the Palermo BAT survey\footnote{http://www.ifc.inaf.it/}. 

In the past decades, the \textit{International Gamma-Ray Astrophysics Laboratory (INTEGRAL,} \citealt{Winkler03}\textit{)} has monitored hard X-ray sources using the Imager on Board the \textit{INTEGRAL} Satellite (IBIS, \citealt{Ubertini03}). \citet{Bird10} and \citet{Krivonos10} published soft Gamma-ray catalogs and 7-year all-sky hard X-ray source lists.  The \textit{INTEGRAL} mission-based hard X-ray catalogs, which covers energy range between 17 keV and 100 keV, provides a better angular resolution ($\sim12$ arcmin of FWHM, IBIS coded-aperture instrument) than \textit{Swift}-BAT survey catalogs which have 19.5 arcmin of FWHM. However, \textit{INTEGRAL}-IBIS observations are concentrated on the galactic plane and the net average exposures are considerably shorter than those of \textit{Swift}-BAT at high galactic latitudes (figure 1). Outside of the Galactic plane the \textit{Swift}-BAT survey detects more sources of primarily extragalactic nature, such as AGN and clusters. The \textit{INTEGRAL}-IBIS found secure counterparts to 369 AGN from the first 1000 orbits (2002-2010, \citealt{Bird16}) while the \textit{Swift}-BAT 70 month survey catalog identified 872 AGN to BAT-detected sources \citep{Ricci17a}. 

Following the most recent data release of the \textit{Swift}-BAT catalog (70 month\footnote{http://swift.gsfc.nasa.gov/results/bs70mon/}, \citealt{Baumgartner13}), this work extends the \textit{Swift}-BAT survey to 105 months, including observations carried out between 2004 December and 2013 August. This catalog presents the \Nnewtotal\ new detections of hard X-ray sources along with detailed source type classifications.  

In Section 2, we briefly introduce procedures adopted for the catalog generation and identification of counterpart along with comparisons with other catalogs. Section 3 introduces structure of the catalog and describes flux measurements, spectral fit, and monthly light curves. In Section 4, we present the survey sensitivity and discuss the uncertainties on the position of the matched counterparts. Finally, we summarize our results in Section 5.

\section{Procedure}
The data reduction, analysis, and catalog generation of the \textit{Swift}-BAT 105 month survey are conducted following the same procedures as in the  \textit{Swift}-BAT 70 month survey \citep{Baumgartner13}. First, the data are extracted in the eight channel energy bands ($14-20$ keV, $20-24$ keV, $24-35$ keV, $35-50$ keV, $50-75$ keV, $75-100$ keV, $100-150$ keV, and $150-195$ keV) from a single snapshot image. After combining the data into all-sky mosaic images, a total band map images are made from the eight channel bands mosaic images. A blind search for detected sources in the 14-195 keV band images is done based on $4.8\sigma$ detection threshold using {\tt batcelldetect}\footnote{http://heasarc.gsfc.nasa.gov/lheasoft/ftools/ftools\_menu.html}. The task {\tt batcelldetect} is run assuming a PSF FWHM of 19.5 arcmin, a source radius of 15 pixels, a background radius of 100 pixels, and a partial coding threshold of 1\%, following previous \textit{Swift}-BAT publications \citep{Tueller10, Baumgartner13}.  An initial run of  {\tt batcelldetect} is done for source detection.  A separate second run of  {\tt batcelldetect} performs position fitting with a search area of 12 arcmin around the source.  For significant detections above a $4.8\sigma$ detection threshold, we identified their optical counterparts by searching the NED and SIMBAD databases as well as archival X-ray data (e.g., \textit{Swift}-XRT, \textit{Chandra}, \textit{ASCA}, \textit{ROSAT}, \textit{XMM-Newton}, and \textit{NuSTAR}).  

Figure~\ref{fig1} illustrates the sky coverage in the 105 month \textit{Swift}-BAT all-sky hard X-ray survey. \textit{Swift}-BAT observed over 50\% of the sky with more than 15.3 Ms of exposure time, while 90\% of the sky is covered with 11.4 Ms.
The left panel in Figure~\ref{fig1} shows the distribution of exposure times in the survey, and the right panel shows the fraction of the sky covered as a function of exposure time. The middle panel shows an all-sky exposure map in galactic coordinates. By comparison, the 90\% sky coverage of \textit{INTEGRAL} from \citet{Bird16} used is 100 ks and the 50\% is $\approx$600 ks.

\citet{Baumgartner13} corrected the gain shift in the \textit{Swift} mission using the peak offset from 59.5 keV calibration line to determine a gain correction factor. The gain shift has stabilized after the end of 2008, with changes less than $1\%$ throughout the period of this survey analysis (private communication, BAT team). Therefore, for analysis beyond 2009, we adopt the same value of gain shift as the last gain correction in 2008.

\begin{figure*}
\centering
	\includegraphics[width=0.9\linewidth]{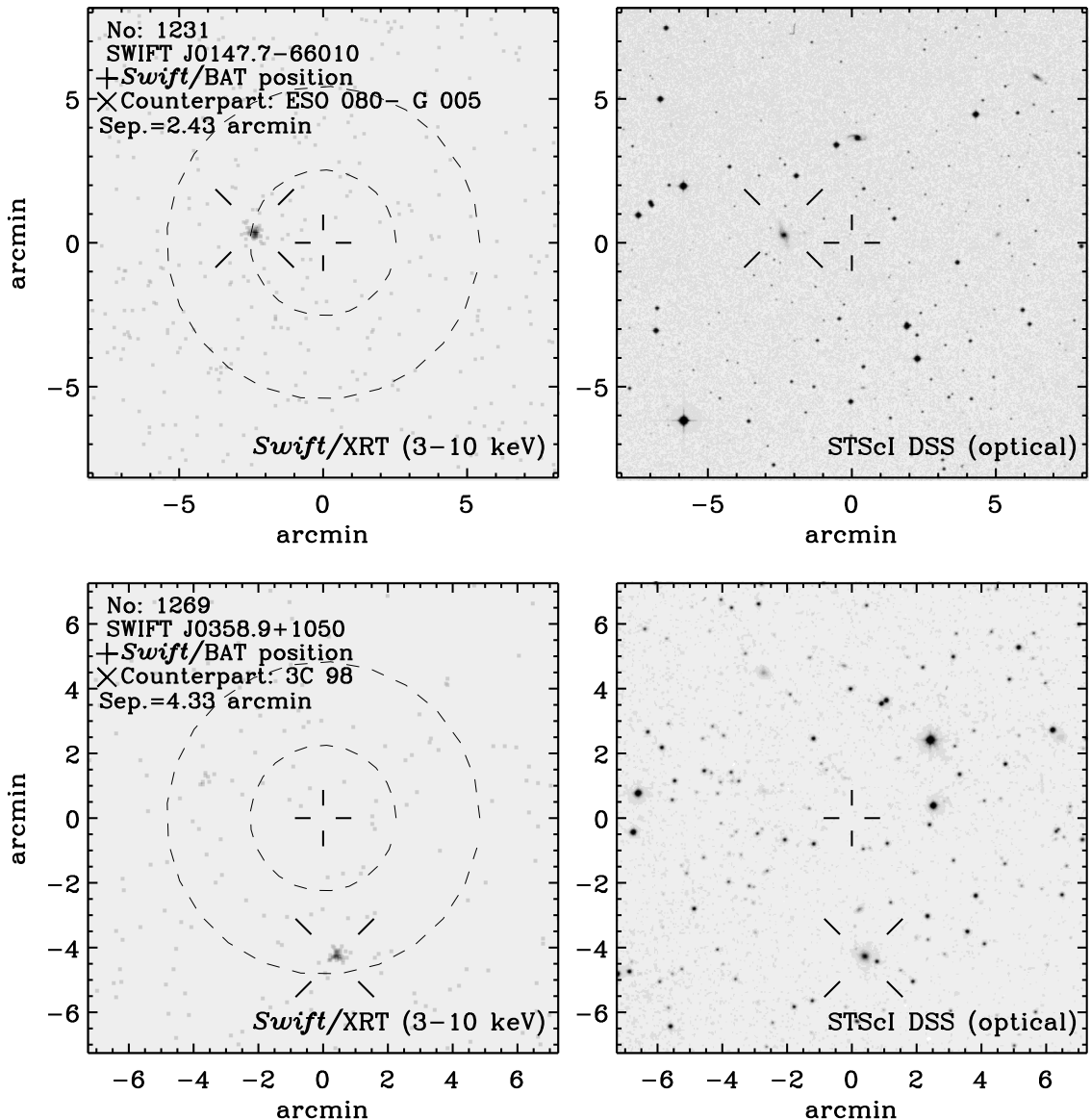}
    \caption{Examples of counterpart identification for ESO 080- G 005 (top panels) and 3C 98 (bottom panels). The \textit{Swift}-BAT source position is shown with a cross hair in the center of each panel. The cross hairs rotated by 45 degrees indicate the position of the identified counterpart. \textit{Left}: The \textit{Swift}-XRT image in $3-10$ keV band. \textit{Right}: The STScI Digitized Sky Survey image. Inner and outer dashed-circles indicate 50\% and 90\% of the \textit{Swift}-BAT positional uncertainty (equation \ref{eq:errequation}), respectively. }    
    \label{ctpt}
\end{figure*}

\subsection{Counterpart identification}
\label{ssec:counterpart}
We inspected soft X-ray images provided by \textit{Swift}-XRT ($3-10$ keV), \textit{Chandra} ($2-10$ keV), \textit{ASCA} ($2-10$ keV) and \textit{XMM-Newton} ($4-10$ keV) for the newly detected sources when available using 15 arcmin of matching radius. Then we compared the soft X-ray images which provide well-defined X-ray coordinates (SNR threshold$>$3) with optical images produced by the STScI Digitized Sky Survey\footnote{https://stdatu.stsci.edu/cgi-bin/dss\_form/}, the Panoramic Survey Telescope and Rapid Response System Data Release 1 (Pan-STARRS DR1\footnote{http://panstarrs.stsci.edu}, \citealt{Chambers16}), the Sloan Digital Sky Survey  Data Release 12 (SDSS DR12\footnote{http://www.sdss.org/dr12/}, \citealt{Alam15}) to confirm their optical counterparts (Figure~\ref{ctpt}). When there are multiple detections in the X-ray images of sources bright enough with the same SNR threshold to be considered as possible counterparts, we assigned `multiple' class, given that the \textit{Swift}-BAT observation has a position uncertainty up to $\approx$13.8 arcmin, which depends on the signal-to-noise ratio (see \ref{ssec:confused}). Based on these procedures, we identified \NnewIdentified\ new hard X-ray sources. The remaining \Nunidentified\ unidentified sources, together with the 35 still unidentified sources listed in the 70 month catalog, are further classified into three categories depending on the presence of soft X-ray observations ($3-10$ keV) and source detections. Sources that have archival soft X-ray observations (e.g., \textit{Swift}-XRT, \textit{Chandra}, and \textit{XMM-Newton}) are assigned to either `Unknown class I' (N=\NUnone) or `Unknown class II' (N=\NUntwo, Table~\ref{tab:types}).  When the observed soft X-ray images do not show well-defined X-ray point source, `Unknown class I' is assigned, and further investigation is required for identification. When there is a soft X-ray detection from archival soft X-ray observation, we assigned `Unknown class II'. We matched these sources whose optical counterpart is not known with Two Micron All Sky Survey (2MASS\footnote{http://irsa.ipac.caltech.edu/Missions/2mass.html}, \citealt{Skrutskie06}) All-Sky Point Source Catalog (PSC) and report the closest counterpart within 5 arcsec of matching radius in Table~\ref{tab:u2}. For the rest `Unknown class III' sources (N=\NUnthree) for which archival X-ray data are not available, we submitted coordinates to the \textit{Swift}-XRT observation with a request of 10 ks of exposure time for a follow-up investigation. Note that we cross-matched all unidentified sources with the archival \textit{NuSTAR} X-ray data and we detected no additional counterparts.

\begin{figure}
\centering
	\includegraphics[angle=90,width=0.5\textwidth]{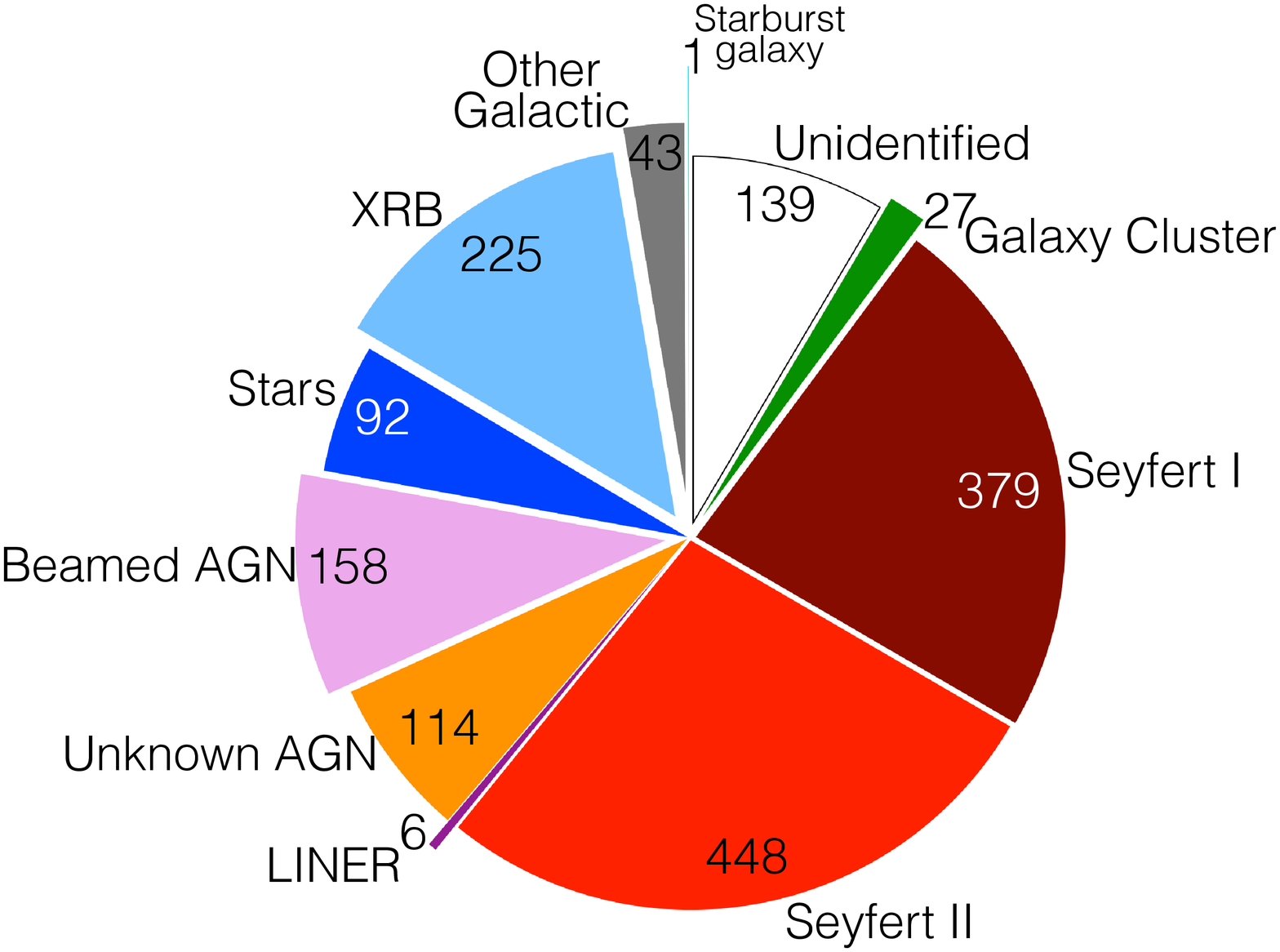}
    \caption{Counterpart source types in the \textit{Swift}-BAT 105 month catalog. 
    `Other Galactic' shown with dark gray wedge includes Galactic center (N=1), pulsars (N=\NPulsar), supernova remnants (N=\NSNR), novae (N=\NNova), globular cluster (N=1), molecular clouds (N=2), gamma-ray source (N=1). `Unidentified' includes Unknown classes (N=\NUnknown) and sources with multiple soft X-ray detections (N=\Nmultiple). `Galaxy Cluster' includes compact group of galaxies (N=1). `Seyfert I' includes high-redshift broad-line AGN (N=2). `Stars' includes cataclysmic variable stars (N=\NCV), symbiotic stars (N=\NSymb), open star cluster (N=1) and other types of stars (N=\Notherstar). `XRB' indicates high mass X-ray binaries (N=\NHMXB), low mass X-ray binaries (N=\NLMXB) and other types of X-ray binaries (N=\NXRB). `Starburst galaxy' indicates M82. See Table~\ref{tab:types} for more detail. (A color version of this figure is available in the online journal.) 
    }
    \label{pie_type}
\end{figure}

\begin{figure*}
\centering
	\includegraphics[width=1\textwidth]{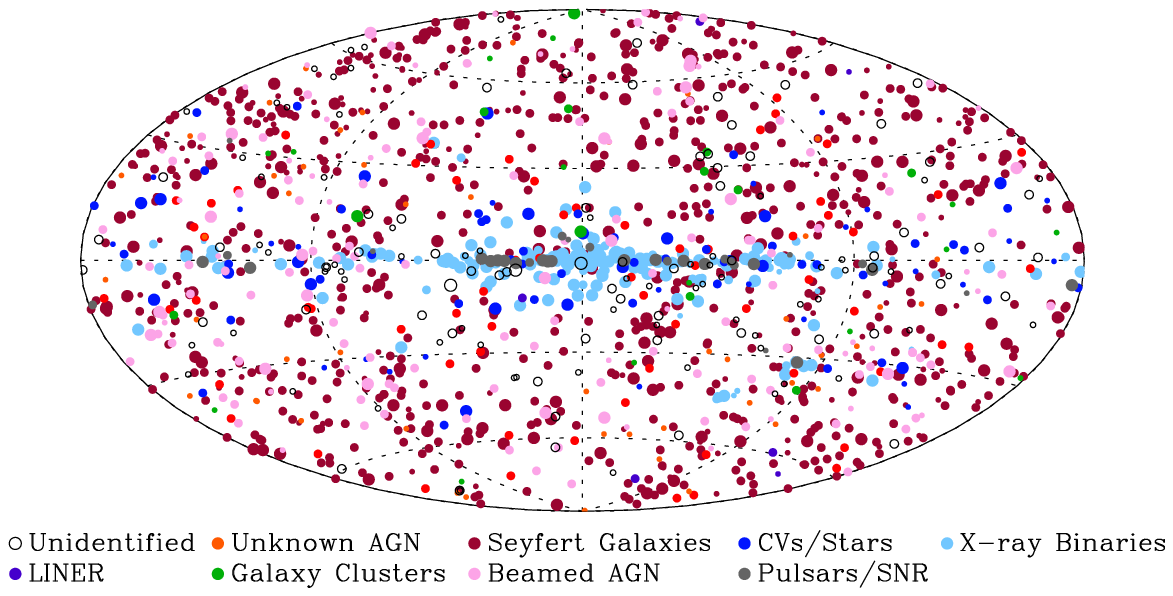}
    \caption{All-sky map showing the classification of the BAT 105 month survey sources. 
    The figure uses a Hammer-Aitoff projection in Galactic coordinates. 
    The size of the filled dots indicate the measured hard X-ray flux in three different scales (log$f_{14-195}<1.0\times10^{-12}$ \ergcms, $1.0\times10^{-12}$ \ergcms$\leq$log$f_{14-195}<1.5\times10^{-12}$ \ergcms, and log$f_{14-195}\geq1.5\times10^{-12}$ \ergcms). 
    The colors correspond to each type of the source as shown in the legend.  
    (A color version of this figure is available in the online journal.) 
    }
    \label{aitoff}
\end{figure*}

\tabletypesize{\scriptsize}
\begin{deluxetable}{p{1.1cm}p{5.0cm}r}
\tablecaption{Counterpart Types in the \textit{Swift}-BAT 105 Month Catalog}
\tablecolumns{3}
\tablehead{
\colhead{Class} &
\colhead{Source Type} &
\colhead{Number} 
}
\startdata
10	&	Unknown class I\tablenotemark{a}			&	\NUnone \\			
11	&	Unknown class II\tablenotemark{b}			&	\NUntwo \\			
12	&	Unknown class III\tablenotemark{c}			&	\NUnthree \\			
15	&	Multiple\tablenotemark{d}					&	\Nmultiple \\		
20	&	Galactic center\tablenotemark{e}			&	1 \\	
30	&	Galaxy Cluster								&	\Ncluster \\			
40	&	Seyfert I (Sy 1.0-1.8)\tablenotemark{f}		&	\NSyone \\			
50	&	Seyfert II (Sy 1.9-2.0)\tablenotemark{g}	&	\NSytwo \\			
60	&	LINER										&	\NLINER \\						
70	&	Unknown AGN\tablenotemark{h}				& 	\NUnknownAGN \\			
80	&	Beamed AGN (Blazar/FSRQ)	 				&	\Nbeamed \\		
90	&	Cataclysmic Variable star (CV)				&	\NCV \\	
100	&	Symbiotic star								&	\NSymb \\			
110	&	Other star									&	\Notherstar \\			
120	&	Open star cluster							&	1 \\	
130	& 	Starburst galaxy\tablenotemark{i}			&	1 \\
140 &   Compact group of galaxies\tablenotemark{j}	&   1 \\		
150	&	Pulsar 										&	\NPulsar \\			
160	&	Supernova remnant (SNR) 					&	\NSNR \\			
170	&	Nova										&	\NNova \\			
180	&	High mass X-ray binary (HMXB) 				&	\NHMXB \\			
190	&	Low mass X-ray binary (LMXB) 				&	\NLMXB \\			
200	&	Other X-ray binary (XRB) 					&	\NXRB \\       		
210	&	Globular Cluster (GC)						& 	1 \\			
220	&	Molecular cloud								&	2 \\			
230	&	Gamma-ray source							&	1 \\			
	&												&	  \\
	&	Total										&	1632
\enddata
\label{tab:types}
\tablenotetext{a}{Unknown class indicates that we do not know source type of the optical counterpart. In particular, `Unknown class I' is used when there is no soft X-ray (3-10 keV) detection despite its archival soft X-ray observation.}
\tablenotetext{b}{`Unknown class II' indicates sources for which there is a soft X-ray detection with SNR threshold greater than 3 from archival soft X-ray observation. When available, we list their 2MASS counterpart source matched within 5 arcsec of angular distance in Table~\ref{tab:u2}.}
\tablenotetext{c}{`Unknown class III' includes sources without archival soft X-ray observations.}
\tablenotetext{d}{`Multiple' class indicates the case where there are more than one soft X-ray detection.}
\tablenotetext{e}{`Galactic center' indicates Sagittarius A*.}
\tablenotetext{f}{`Seyfert I' includes high-z broad-line AGN (N=2).}
\tablenotetext{g}{`Seyfert II' includes Sy2 candidates (N=10) which present narrow emission lines in their optical spectral energy distribution.}
\tablenotetext{h}{`Unknown AGN' indicates X-ray sources associated with galaxies whose optical spectra and type classifications are not known.}
\tablenotetext{i}{`Starburst galaxy' indicates M82.}
\tablenotetext{j}{`Compact group of galaxies' indicates Arp 318.}
\end{deluxetable}

\subsection{Source type}
\label{ssec:type}
Since we have reclassified source types (Table~\ref{tab:types}) compared to that of the previous \textit{Swift}-BAT 70 month survey catalog, we reviewed the classification of sources reported by the 70 month survey catalog to properly assign their source types following the current scheme. The main changes in counterpart source types compared to that of 70 month survey catalog are in the  AGN (including `Unknown AGN', `LINER', `Other AGN', and `Seyfert' classes) `Galaxy', `QSO', and `Unknown' classes. We assigned `Unknown AGN' when a source is associated with an extended galaxy in the optical image but lacks firm evidence of AGN optical emission diagnostics. `QSO' type is replaced with either one of sub-classes of `Seyfert' or `Beamed AGN (Blazar/FSRQ)' based on the presence and shape of optical emission lines in the literature. 

In order to provide reliable classification of source type, we searched public optical spectroscopic surveys (Sloan Digital Sky Survey and 6dF Galaxy Survey; \citealt{Abazajian09, Jones09, Alam15}), the follow-up improved spectral line measurement database of SDSS DR7 galaxies (the OSSY catalog, \citealt{Oh11}\footnote{http://gem.yonsei.ac.kr/ossy/}), the 13th edition of quasars and AGN catalogues (\citealt{VeronCetty10}\footnote{http://cdsweb.u-strasbg.fr/cgi-bin/qcat?J/A+A/518/A10/}) as well as recent investigation from the \textit{Swift}-BAT AGN Spectroscopic Survey \citep{Koss17, Ricci17a} and \textit{INTEGRAL} mission based X-ray catalogues (\textit{INTEGRAL} General Reference catalog\footnote{v40, http://www.isdc.unige.ch/integral/science/catalogue\#Reference}, \textit{INTEGRAL} IBIS/ISGRI catalog, \citealt{Beckmann09, Malizia16}) to determine detailed source types that listed in Table~\ref{tab:types}. We also used the Roma blazar catalogue (BZCAT) v5.0\footnote{http://www.asdc.asi.it/bzcat/} \citep{Massaro09} for reference to beamed AGN. 

Table~\ref{tab:types} and Figure~\ref{pie_type} give a summary of counterpart identifications for the \textit{Swift}-BAT 105 month catalog. Figure~\ref{aitoff} presents the distribution of sources in a Galactic coordinates with a Hammer-Aitoff projection. The colors indicate the different types of sources and the size of the symbol corresponds to the source flux in the $14-195$ keV band. 

\begin{figure*}
\centering
	\includegraphics[angle=270,width=0.8\textwidth]{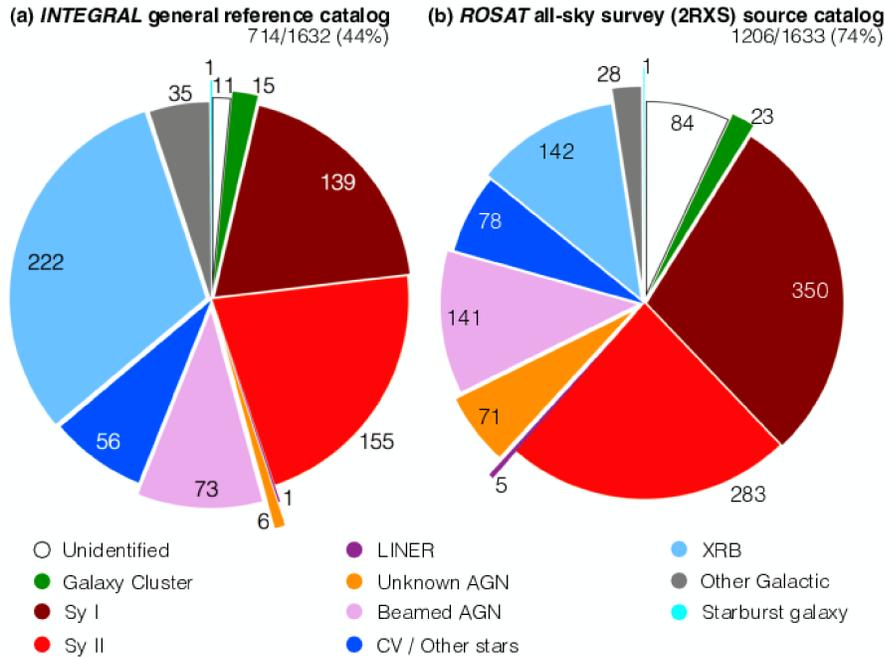}
    \caption{Counterpart types in common with the \textit{Swift}-BAT 105 month sources. \textit{Left:} \textit{INTEGRAL} general reference catalog. \textit{Right:} The second \textit{ROSAT} all-sky survey (2RXS) source catalog. Number of sources is shown in the pie chart. Note that the majority of sources in `CV / Other stars' category come from CV (see Table~\ref{tab:types}). (A color version of this figure is available in the online journal.) 
    }
    \label{cross_matched_catalog}
\end{figure*}

\begin{figure*}
\centering
	\includegraphics[angle=90,width=0.8\textwidth]{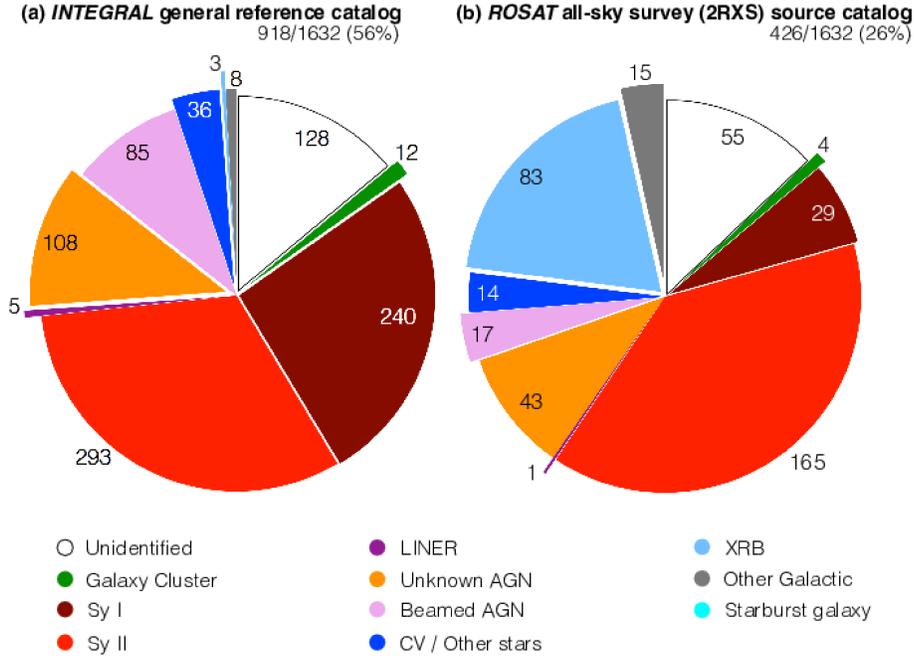}
    \caption{\textit{Swift}-BAT 105 month sources that were not reported in the \textit{INTEGRAL} general reference catalog (\textit{left}) and the  second \textit{ROSAT} all-sky survey (2RXS) source catalog (\textit{right}). The format is same as that of Figure~\ref{cross_matched_catalog}. (A color version of this figure is available in the online journal.)  
   }
    \label{not_matched_catalog}
\end{figure*}

Figure~\ref{cross_matched_catalog} shows the counterpart types of the \textit{Swift}-BAT 105 month sources in common with the \textit{INTEGRAL} general reference catalog (\textit{left}) and the \textit{ROSAT} all-sky survey (2RXS) source catalog (\textit{right}, \citealt{Boller16}). As the \textit{ROSAT} all-sky survey scanned the whole sky in the $0.1-2.4$ keV band \citep{Truemper82}, a large fraction of the overlap with BAT and \textit{ROSAT} is found for sources identified as Seyfert I due to absence of absorption in these objects. On the other hand, the \textit{INTEGRAL} catalog has a large overlap with the BAT for galactic sources, such as X-ray binaries, due to its deep galactic plane exposures. In Figure~\ref{not_matched_catalog}, we present the number of each sources of the various types that are not detected in the \textit{INTEGRAL} general reference catalog (\textit{left}) and the \textit{ROSAT} all-sky survey (2RXS) source catalog (\textit{right}).  It is clearly seen that \textit{Swift}-BAT detects a large number of extragalactic sources, primarily AGN, that are not detected by either \textit{ROSAT} or \textit{INTEGRAL}.

The Palermo \textit{Swift}-BAT hard X-ray catalog identified 1286 hard X-ray sources in their latest publication \citep{Cusumano10b} using data acquired in the 54 months of the \textit{Swift}-BAT operation. The catalog adopted a $4.8\sigma$ of significance threshold from at least one of the three energy bands ($15-30$ keV, $15-70$ keV, and $15-150$ keV). The total number of spurious detections varies between 15 and 45 depending on detection in each energy band, compared to one spurious detection in the entire sky at the same threshold level achieved by the BAT catalogs as \citet{Tueller10} described. 

We measure the overlap between the Palermo 54 month BAT catalog and this work using a 15 arcmin matching radius to test whether the BAT detection maps agree between the two processing routines. Since the 54 month catalog was released in 2010 before much of the soft X-ray data was available, 30 out of 1286 sources have multiple counterparts listed for a single BAT detection, which we exclude. We confirmed that \fracbatmatchedwithPalermo\% of the sources identified in this work (\NBATmatchedwithPalermo/\Ntotal) are in overlap with the 54 month Palermo \textit{Swift}-BAT catalog. Of the \NBATmatchedwithPalermo\ sources common in both catalogs, the 105 month catalog presents \NBATnewmatchedwithPalermo\ new detections with respect to the 70 month catalog. The majority of the common sources (\NAGNXRBmatchedwithPalermo/\NBATmatchedwithPalermo) are AGN (N=\NAGNmatchedwithPalermo) or X-ray binaries (N=\NXRBmatchedwithPalermo). Meanwhile, \NexclusivelyinBAT\ sources are found in the \textit{Swift}-BAT 105 month catalog but not in the Palermo 54 month catalog. Among those sources not detected in the Palermo 54 catalog, bulk populations are AGN (N=\NAGNexclusivelyinBAT) or unknown type of sources (N=\NunknownexclusivelyinBAT). It is also noteworthy to mention that the \NexclusivelyinPalermo\ sources are detected in the 54 month Palermo \textit{Swift}-BAT catalog but not in this work. Of the \NexclusivelyinPalermo\ sources detected in Palermo but not detected in the 105 month catalog at $14-195$ keV, the majority are detected at softer energies either in the $15-30$ or $15-70$ keV band (\fracsofterband\%, \Nsofterband/\NexclusivelyinPalermo). The \NexclusivelyinPalermo\ sources that are only detected in the 54 month Palermo \textit{Swift}-BAT catalog will be investigated further in future catalogs that use alternate detection strategies with many energy bands, time variability, or are optimized for sources like heavily obscured AGN (e.g. \citealt{Koss16}). Future catalog papers will compare sources in the Palermo 100 month catalog (Segreto et al. in prep).

\begin{figure*}[ht]
	\includegraphics[width=\linewidth]{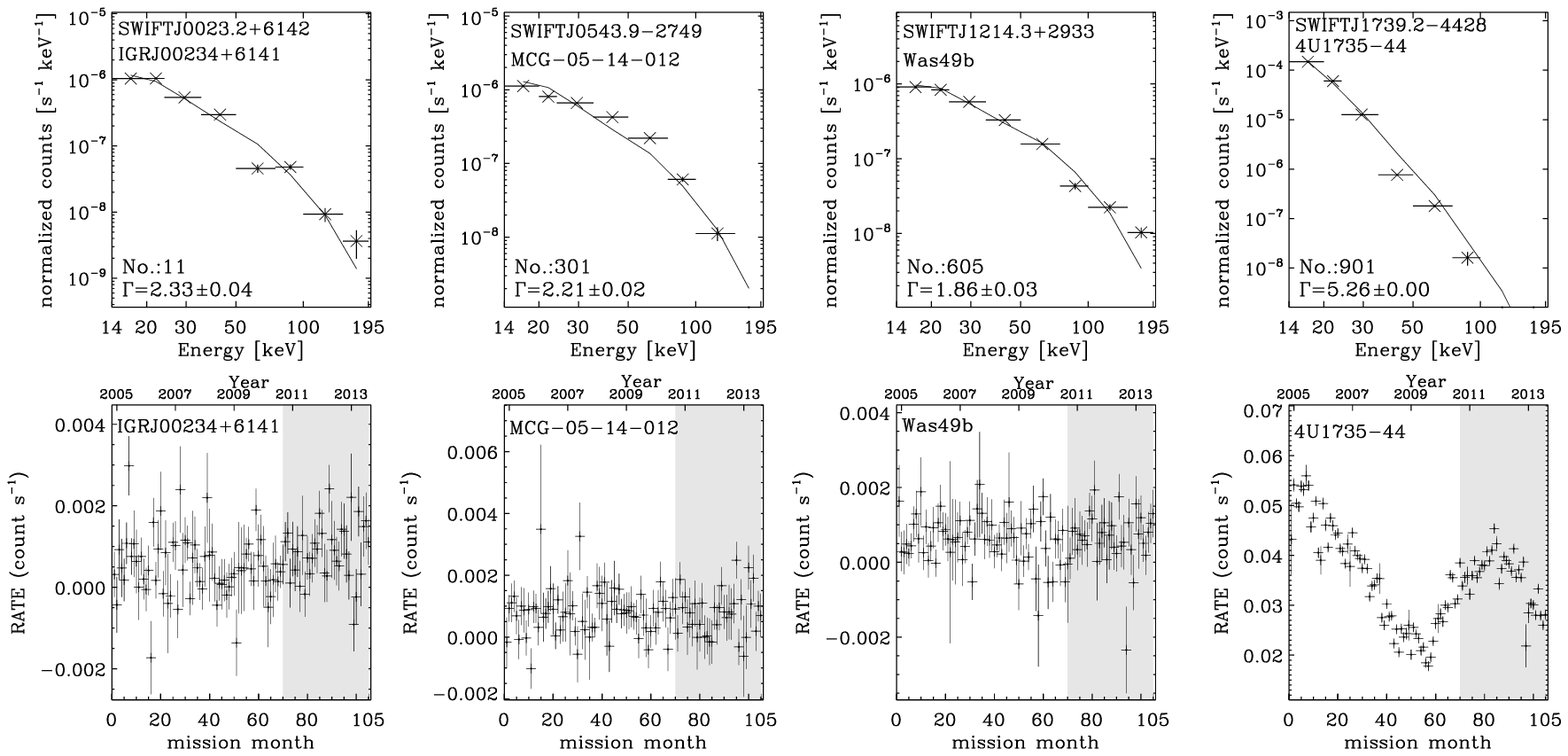}
    \caption{Example spectra and fits (\textit{top} panels), light curves (\textit{bottom} panels) of four \textit{Swift}-BAT sources in the 105 month catalog. 
    \textit{Top:} Normalized counts in unit of ${\rm s^{-1}\ keV^{-1}}$ for the eight-channel are shown with solid lines of a simple power-law fit. The \textit{Swift}-BAT name, counterpart name, BAT number, and spectral photon index ($\Gamma$) are shown in the legend.     
    \textit{Bottom:} Grey shades indicate the observation period from the end of 70th month (September 2010) to the end of 105th month (August 2013). 
    (The complete set is available in the \textit{Swift}-BAT 105 month Web site.\textsuperscript{\ref{web}})} 
    \label{spec_and_lc}
\end{figure*}

\section{The \textit{Swift}-BAT 105 month catalog}
\label{sec:catalog}
Table~\ref{tab:catalog} presents the catalog of the \textit{Swift}-BAT 105 month data. The 105 month catalog in its full extent can be found in the online version of the journal, and on the \textit{Swift}-BAT 105 month survey Web site\footnote{http://swift.gsfc.nasa.gov/docs/swift/results/bs105mon/\label{web}}. Future multiwavelength counterpart data for the \textit{Swift}-BAT 105 month sources will also be available\footnote{www.bass-survey.com}. In Table~\ref{tab:catalog} we present the source number in the first column, which has a consistent order with the previous \textit{Swift}-BAT 70 month survey catalog. We listed the sources reported from the \textit{Swift}-BAT 70 month survey catalog first ($1-1210$), and then we assigned following numbers ($1211-1632$) for the newly detected sources in the order of increasing right ascension.  

\begin{figure*}
\centering
	\includegraphics[width=0.97\linewidth]{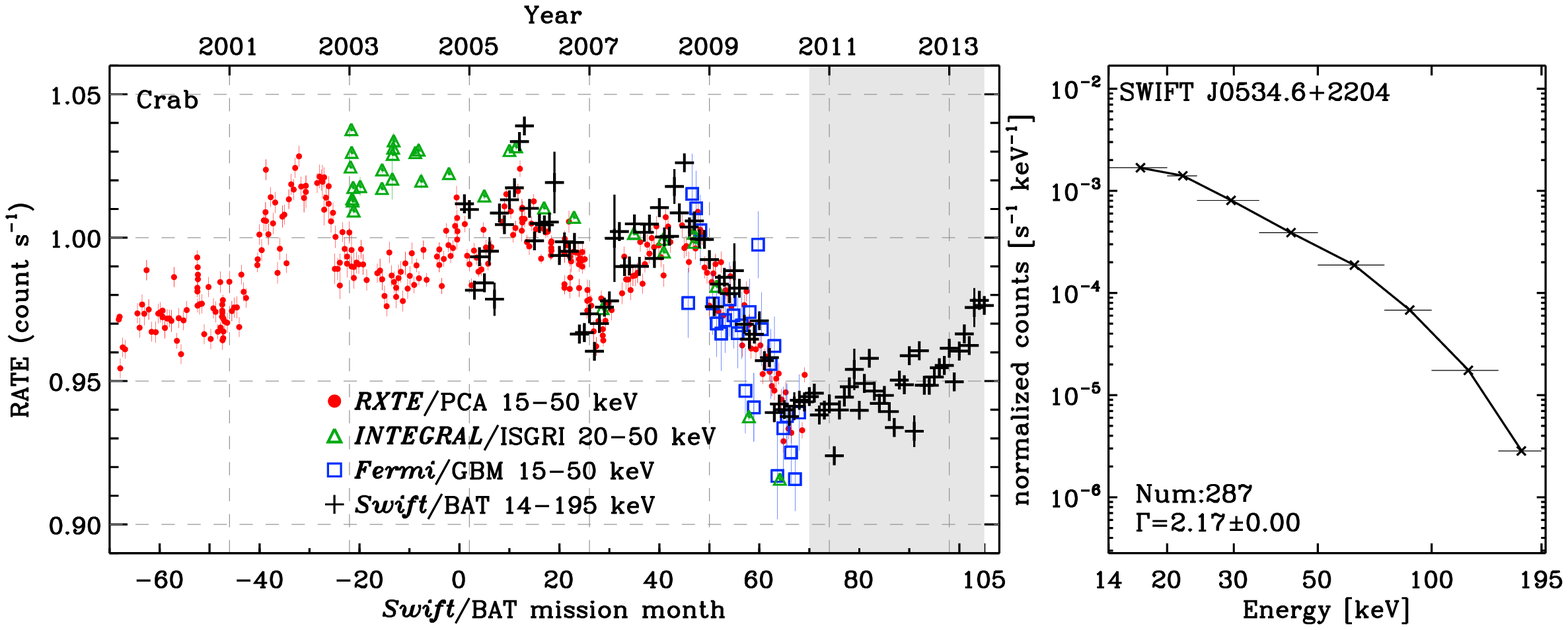}
    \caption{Composite light curves (\textit{left} panel) and a spectrum with a fit (\textit{right} panel) of the Crab nebula. \textit{RXTE}/PCA ($15-50$ keV: red filled dots), \textit{INTEGRAL}/ISGRI ($20-50$ keV: open green triangles), \textit{Fermi}/GBM ($15-50$ keV: open blue squares), and \textit{Swift}/BAT ($14-195$ keV: black crosses) data are shown based on the \textit{Swift}/BAT mission month. Data sets have been normalized to its mean rate in the observation period from August to November 2008. The grey shaded area indicates the observation period from September 2010 to August 2013. Right panel shows the spectral fit of Crab nebula with its \textit{Swift} name, catalog number and the measured spectral photon index. (A color version of this figure is available in the online journal.)}
    \label{Crab_LC_spectrum}
\end{figure*}

We provide the BAT name in the second column, which is created by the BAT source position along with the \textit{Swift}-BAT detection in equatorial coordinates for epoch J2000.0 in the third and the fourth columns. For the sources which has been previously published (i.e., the 70 month \textit{Swift}-BAT catalog), we have used the corresponding BAT name.  

The significance of the BAT source detection which is the ratio between the highest pixel value of the source from the total-band mosaic flux map and the local noise is listed in the fifth column. As discussed in \citet{Tueller10}, we chose $4.8\sigma$ significance level as a detection threshold since one can expect 1.54 sources at the $4.8\sigma$ level from $1.99\times10^{6}$ independent pixels. 

We also provide the name of the counterpart in the sixth column, identified as described in Section~\ref{ssec:counterpart}. The name of the counterpart can be a well-known optical source, or it can be a wavelength dependent. Often these names come from the soft band X-ray archival data such as \textit{Swift}-XRT, \textit{Chandra}, and \textit{XMM-Newton}. For the firmly identified counterpart sources, we provide their equatorial coordinates (J2000.0) in the seventh and eighth columns. 

Table~\ref{tab:catalog} includes $14-195$ keV flux of the \textit{Swift}-BAT sources in units of ${\rm 10^{-12}\ erg\ s^{-1}\ cm^{-2}}$, with its 90\% confidence level in the ninth and tenth columns. The \textit{Swift}-BAT flux of each counterpart is extracted from the hard X-ray map at the location of the identified counterpart given in the seventh and eighth columns. In order to measure flux and photon spectral index ($\Gamma$), a power-law model is applied. 

The \textit{Swift}-BAT luminosity of the counterpart is also provided in Table~\ref{tab:catalog} for sources classified as AGN. We assumed a cosmology with $H_{\rm 0}= 70 {\rm km\ s^{-1}\ Mpc^{-1}}$, $\Omega_{\rm m}=0.30$, and $\Omega_{\rm \Lambda}=0.70$ for computing the source luminosity in a unit of \ergs\ in the $14-195$ keV band using the redshift and flux listed in the table.

\tabletypesize{\scriptsize}
\begin{deluxetable*}{p{0.4cm}lrrc lrrr}[ht]
\tablecaption{Unidentified sources with 2MASS counterpart in the \textit{Swift}-BAT 105 Month Catalog}
\tablecolumns{9}
\tablehead{
\colhead{Num} &
\colhead{BAT Name} &
\colhead{R.A.\tablenotemark{a}} &
\colhead{Decl.\tablenotemark{a}} & 
\colhead{Source Type\tablenotemark{b}} &
\colhead{2MASS designation} &
\colhead{R.A.\tablenotemark{c}} &
\colhead{Decl.\tablenotemark{c}} &
\colhead{Sep.\tablenotemark{d}}
}
\startdata
1293&                SWIFT J$0628.7-8346$&                97.4441667&             $-83.7394722$&                   U2&                  $06294867-8344217$&                 97.452801&              $-83.739380$&            3.40 \\
1340&                SWIFT J$0850.8-4219$&               132.6658333&             $-42.1977500$&                   U2&                  $08504008-4211514$&                132.667015&              $-42.197617$&            3.20 \\
1362&                SWIFT J$0958.2-5732$&               149.6466667&             $-57.4901389$&                   U2&                  $09583496-5729199$&                149.645688&              $-57.488884$&            4.90 \\
1454&                SWIFT J$1503.7-6028$&               226.0675000&             $-60.3559444$&                   U2&                  $15041611-6021225$&                226.067125&              $-60.356274$&            1.38 \\
1479&                SWIFT J$1617.9-5403$&               244.5320833&             $-54.1036944$&                   U2&                  $16180771-5406122$&                244.532140&              $-54.103394$&            1.07 \\
1525&                SWIFT J$1800.8-4148$&               270.1758333&             $-41.7804722$&                   U2&                  $18004247-4146466$&                270.176973&              $-41.779613$&            4.35 \\
1554&                SWIFT J$1857.6-0748$&               284.3962500&              $-7.5314167$&                   U2&                  $18573532-0731513$&                284.397176&               $-7.530917$&            3.77 \\
1583&                SWIFT J$2033.1+0991$&               308.4137500&               $9.8260278$&                   U2&                  $20333946+0949338$&                308.414420&                $9.826065$&            2.38 \\
1591&                SWIFT J$2047.0+4112$&               311.6637500&              $41.3304722$&                   U2&                  $20463963+4119471$&                311.665153&               $41.329754$&            4.59 \\
1592&                SWIFT J$2055.0+3559$&               313.7841667&              $35.9416389$&                   U2&                  $20550835+3556278$&                313.784820&              $313.784820$&            2.78 \enddata
\label{tab:u2}
\tablenotetext{a}{The \textit{Swift}-BAT XRT soft X-ray (3-10 keV) source position in J2000 coordinates.}
\tablenotetext{b}{U2 indicates `Unknown class II' which is listed in the Table.~\ref{tab:types}.}
\tablenotetext{c}{The 2MASS source position in J2000 coordinates.}
\tablenotetext{d}{Angular separation between the \textit{Swift}-XRT soft X-ray source position and 2MASS catalog in unit of arcsecond.}
\end{deluxetable*}

Lastly, Source type class shown in Table~\ref{tab:types} in integer is provided with detailed source type in the last two columns.

\begin{figure}
	\includegraphics[width=\linewidth]{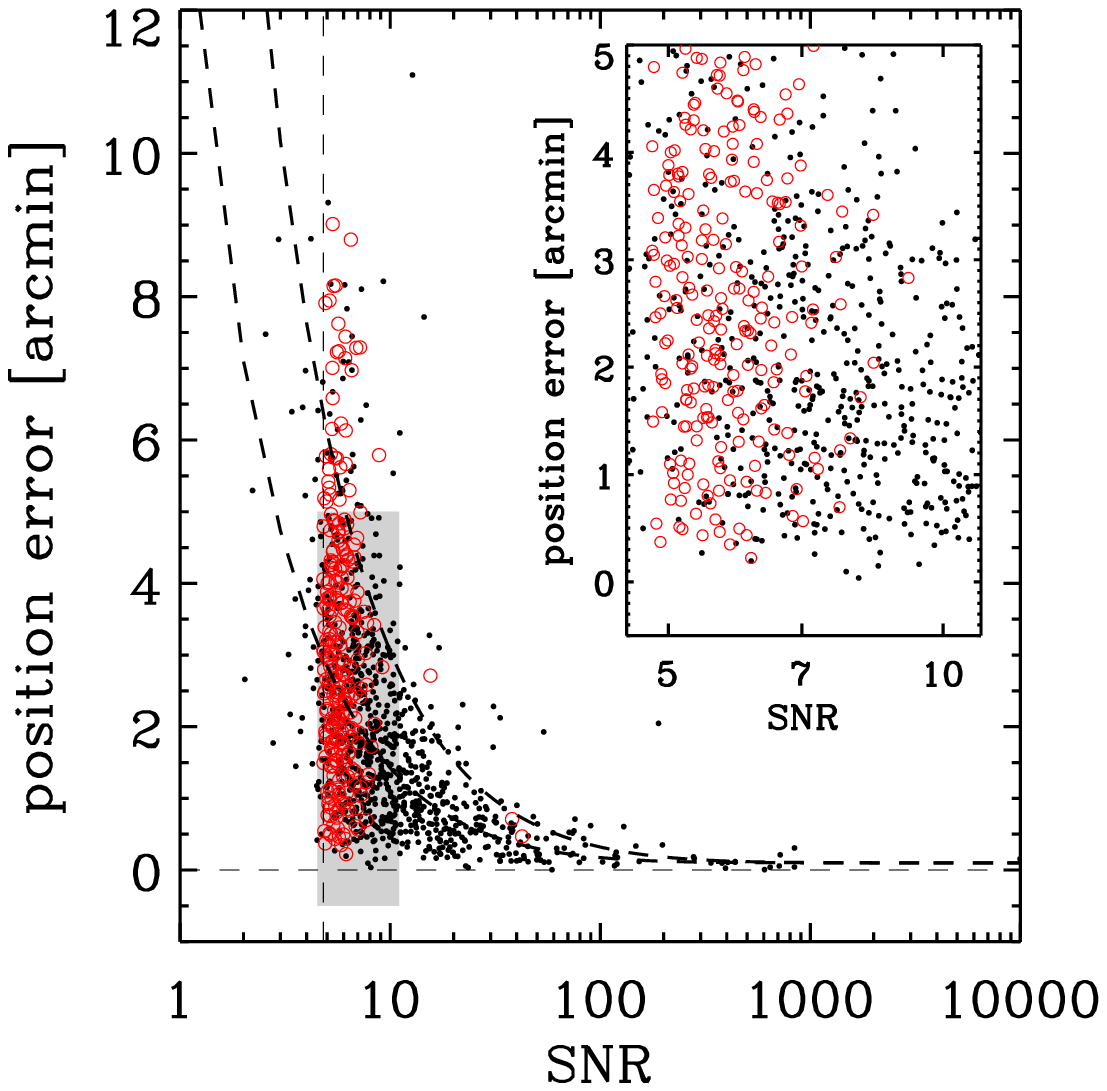}
    \caption{The \textit{Swift}-BAT position error in unit of arcminute as a function of the detection significance. Sources closer than 5 degree from the galactic plane are not included. Sources identified from the \textit{Swift}-BAT 70 month survey are shown with black filled dots, while the new detections from the \textit{Swift}-BAT 105 month survey are shown with red open circles. The two dashed curves show the 90\% and 50\% error radius as a function of detection significance. The inset panel, which corresponds to the grey area in the main panel, shows the low SNR regime. (A color version of this figure is available in the online journal.)}
    \label{position_err}
\end{figure}

\subsection{BAT Fluxes and Spectra}
\label{ssec:flux}
The fluxes listed in the Table~\ref{tab:catalog} were measured from the eight BAT bands ($14-20$ keV, $20-24$ keV, $24-35$ keV, $35-50$ keV, $50-75$ keV, $75-100$ keV, $100-150$ keV, and $150-195$ keV) of the mosaicked maps at the position of the identified counterpart. For the case of BAT sources whose counterparts are not known, we measured the flux from the detected BAT source position using the same eight BAT bands. 

We measured the fluxes of the \textit{Swift}-BAT sources by fitting the eight-channel spectra with the {\tt pegpwrlw} model (power-law with pegged normalization) provided by the {\tt XSPEC}\footnote{http://heasarc.nasa.gov/xanadu/xspec/} software \citep{Arnaud96} over the $14-195$ keV range as \citet{Tueller10} and \citet{Baumgartner13} presented in previous \textit{Swift}-BAT catalogs. However, the applied {\tt pegpwrlw} model does not always yields a good fit for the \Ntotal\ \textit{Swift}-BAT sources, since a simple power-law model cannot explain the wide range of physical properties of the different objects. For such reason, we also provide the reduced $\chi^{2}$ value for each source in Table~\ref{tab:catalog} as an indicator of goodness of fit. 

We also used the {\tt error} function provided by {\tt XSPEC} in order to estimate error values of the overall flux and the spectral photon index with the 90\% confidence interval. 

We provide the eight-channel spectra of the \textit{Swift}-BAT 105 month survey sources in a format of standard fits file on our Web site\textsuperscript{\ref{web}} and in the online journal. Examples of the spectra and their fits are shown in the Figure~\ref{spec_and_lc} (top panels).

\begin{figure*}
\centering
	\includegraphics[width=0.9\linewidth]{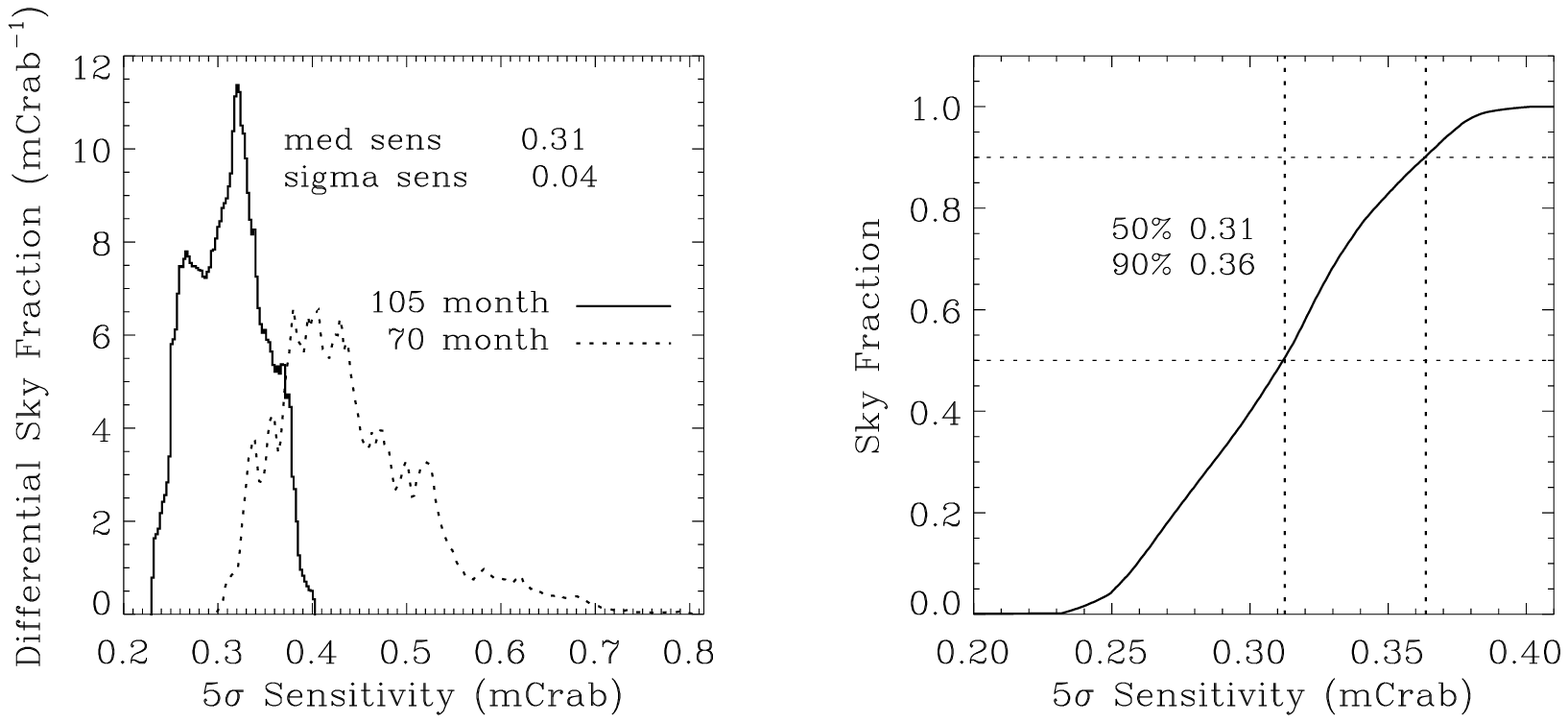}
    \caption{Sky coverage vs. sensitivity. \textit{Left:} Solid line and dotted line represent differential sky fraction for the 105 month and the 70 month survey, respectively. \textit{Right:} The 0.31 mCrab sensitivity limit of the 105 month survey covering 50\% of the sky corresponds to a flux of \senshalf.}
    \label{sens}
\end{figure*}

\begin{figure}
	\includegraphics[width=\linewidth]{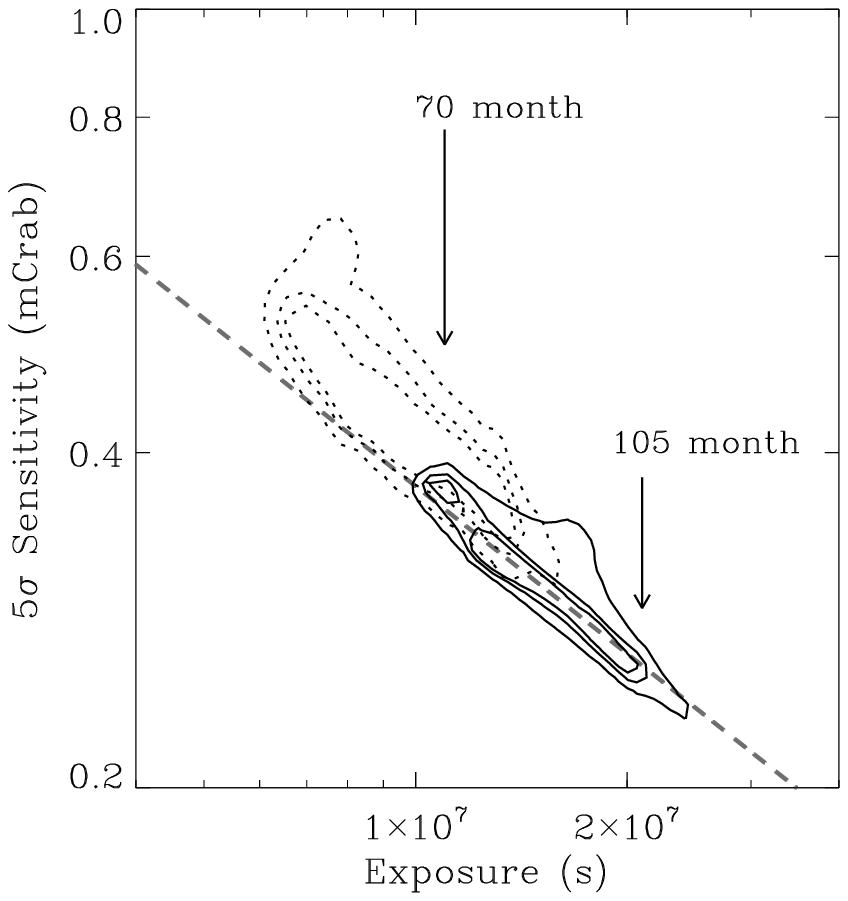}
    \caption{The $5\sigma$ sensitivity limit for pixels in the all-sky map as a function of effective exposure time for the 70 month (dotted contours), and 105 month survey (solid contours). The dashed line represents a lower limit to the expected Poisson noise.}
    \label{sens_exp}
\end{figure}

\subsection{Light Curves}
\label{ssec:lightcurves}
Similarly to what was done in the \textit{Swift}-BAT 70 month catalog \citep{Baumgartner13}, we provide monthly light curves of the \textit{Swift}-BAT sources that span the 105 month period between December 2004 and August 2013. For each month of \textit{Swift}-BAT data, we generated all-sky total-band mosaic images. Then we extracted the monthly mosaic fluxes for each hard X-ray detection that identified over the whole period of observation. Figure~\ref{spec_and_lc} shows four example light curves from the \textit{Swift}-BAT 105 month survey in the bottom panels.

The Crab nebula is of particular importance as it has been used as a standard candle by X-ray and gamma-ray studies. Our results for the Crab nebula are consistent with those reported \citet{Wilson-Hodge11} with a $\sim7\%$ decline in the $15-50$ keV band from the observation of the \textit{Fermi} Gamma-ray Burst Monitor \citep{Meegan09}. Figure~\ref{Crab_LC_spectrum} shows composite light curves of the Crab nebula from \textit{RXTE}, \textit{INTEGRAL}, \textit{Fermi}, and \textit{Swift}. The \textit{Swift}-BAT 70 month data (\textit{Swift}/BAT mission month from 1 to 70 in the left panel of Figure~\ref{Crab_LC_spectrum}) confirmed the variations of 10\% in a long-term scales up to the first 70 month of observations, i.e., September 2010. As illustrated in the left panel of Figure~\ref{Crab_LC_spectrum}, the monthly light curve of the Crab nebula increases since September 2010 ($\sim4\%$). 

We provide the light curves of the \textit{Swift}-BAT 105 month survey sources on the \textit{Swift}-BAT Web site\textsuperscript{\ref{web}} and in the online journal along with the eight-channel spectra.

\subsection{Confused Sources}
\label{ssec:confused}
As mentioned in the Section~\ref{ssec:counterpart} and Table~\ref{tab:types}, \Nmultiple\ sources are classified as `multiple' in our catalog when there are at least two detections in \textit{Swift}-XRT soft X-ray image. In these cases, it is possible that the detected \textit{Swift}-BAT source has multiple X-ray counterparts significantly contributing to the observed $14-195$ keV flux. We explicitly differentiate those sources in Table~\ref{tab:types} and Table~\ref{tab:catalog} using distinctive source class ($15$) and type (`Multiple').  

In the previous 70 month catalog multiple bright soft X-ray counterparts were found and they were all listed because of the inability of the low quality \textit{Swift} XRT to enable detailed spectral modeling to find the likely sources. However, subsequent work found that the BAT emission was often dominated by a single AGN except in a few unusual cases  (e.g. \citealt{Ricci17a}). For instance the typical luminosity ratio for dual AGN is large (1/11, \citealt{Koss12}). In this catalog version we report only one entry for each BAT source with the `multiple' counterpart flag and will discuss the likely counterparts in future papers when better X-ray data is available.

\section{Survey Characteristics}
\label{sec:characteristics}

\subsection{Source Positions and Uncertainties}
\label{ssec:uncertainties}
We show the angular separation between the \textit{Swift}-BAT position and the position of identified counterpart as a function of the level of the significance of the \textit{Swift}-BAT source detection in Figure~\ref{position_err}. As one can expect, the angular separation (i.e., the positional accuracy of the \textit{Swift}-BAT detection) decreases as the level of the detection significance becomes larger. 

For the identified counterparts which are not located in the Galactic plane (Galactic latitude $>+5$ or $<-5$ degree), the estimate of the error radius (in arcminutes) can be represented as follows
\begin{equation}
\text{BAT error radius (arcmin)} = \sqrt{{\Big(\frac{30.5}{\text{SNR}}\Big)}^{2} + 0.1^{2}}
	\label{eq:errequation}
\end{equation}
where SNR is the level of the significance of the \textit{Swift}-BAT detection. This empirical function explains the position error radius for the 90\% of the \textit{Swift}-BAT sources that are not in the galactic plane. The systematic error of 0.1 arcmin is deduced from the positional error of high SNR sources ($>200$).

\subsection{Measured Sensitivity and Noise}
\label{ssec:sens_noise}
Figure~\ref{sens} presents the distribution of $5\sigma$ sensitivity in units of mCrab measured from the all-sky mosaicked map. The achieved median $5\sigma$ sensitivity in the 105 month survey is 0.31 mCrab, which corresponds to \senshalf\ in the $14-195$ keV band, while 90\% of the sky is covered with a sensitivity of 0.36 mCrab (\sensninety)\footnote{A total Crab flux in the $14-195$ keV band is $2.3343\times10^{-8}$ \ergcms}.

Figure~\ref{sens_exp} shows the measured sensitivity in unit of mCrab as a function of exposure time. The solid contours are from the 105 month survey while the dotted contours are from the 70 month survey as a comparison. As \citet{Baumgartner13} described, the curved tails shown in the 70 month survey in the short exposure and high $5\sigma$ sensitivity regime arise from observations near galactic center with high systematic noise. The 105 month data are consistent with a sensitivity that scales with the square root of the exposure time, indicating that systematic limits to the BAT sensitivity have not yet been reached.

\section{Conclusions}
\label{sec:conclusion}
The \textit{Swift}-BAT 105 month catalog is the fifth and the most recent catalog of the \textit{Swift}-BAT all-sky hard X-ray survey. 
The catalog includes \Ntotal\ hard X-ray sources detected across the entire sky. Out of these \Ntotal\ sources, \Nnewtotal\ are new detections with respect to the 70 month catalog, and 320 are reported as hard X-ray sources for the first time. The \textit{Swift}-BAT 105 month survey catalog contains \NtotAGN\ non-beamed AGN detected in the hard X-ray band. Given the fact that most of the \Nunknown\ unidentified sources are not located near the galactic plane, the total number of AGN is likely be more than the current report. Furthermore, the catalog includes list of beamed AGN (blazars/FSRQ), cataclysmic variable star (CV), pulsar, supernova remnant, high mass X-ray binary (HMXB) as well as low mass X-ray binary (LMXB) which are important references for many scientific studies. 

Follow-up observations and studies of the \textit{Swift}-BAT sources are being actively pushed forward. \citet{Berney15} showed correlations between X-ray continuum emission and optical narrow emission lines based on the optical spectroscopic follow-up project (the BAT AGN Spectroscopic Survey, \citealt{Koss17}). The relationship between optical narrow-line emission line ratios and Eddington accretion rate was investigated by \citet{Oh17}. \citet{Lamperti17} also explored near-Infrared (NIR, $0.8-2.4\micron$) spectroscopic properties of 102 \textit{Swift}-BAT selected AGN. Detailed studies of X-ray properties such as spectral photon index \citep{Trakhtenbrot17} and obscuration \citep{Ricci17b} have been made, and the correlation between merger stage of interacting systems and their Eddington accretion rate (Koss et al., in prep.) is currently in preparation. Meanwhile, \textit{Swift}-XRT observations are currently being carried with 10 ksec of exposure time as part of a filler program for all the sources detected by \textit{Swift}-BAT without soft X-ray observations.

Time variability, soft and hard bands studies of the \textit{Swift}-BAT sources will be explored in future catalogs. Furthermore, while we selected sources based on a 4.8$\sigma$ detection threshold using a Crab like spectra, more sources can be detected using the spectral curvature or harder/softer $\Gamma$ \citep{Koss13, Koss16} which we also plan to publish in future catalogs. As a final remark, we address that we are currently working on the 148 months catalog as of time of this writing.

\section*{Acknowledgments}
KO and KS acknowledge support from the Swiss National Science Foundation (SNSF) through Project grants 200021\_157021. KO is an International Research Fellow of the Japan Society for the Promotion of Science (JSPS) (ID: P17321). MK acknowledges support from the SNSF through the Ambizione fellowship grant PZ00P2\_154799/1, SNSF grant PP00P2 138979/1, and support from NASA through ADAP award NNH16CT03C.
KS acknowledges support from Swiss National Science Foundation Grants PP00P2\_138979 and PP00P2\_166159. 

This research has made use of data and/or software provided by the High Energy Astrophysics Science Archive Research Center (HEASARC), which is a service of the Astrophysics Science Division at NASA/GSFC and the High Energy Astrophysics Division of the Smithsonian Astrophysical Observatory. 

This research has made use of data supplied by the UK Swift Science Data Centre at the University of Leicester.

The scientific results reported in this article are based on observations made by the Chandra X-ray Observatory, data obtained from the Chandra Data Archive, observations made by the Chandra X-ray Observatory and published previously in cited articles.

This research has made use of observations obtained with XMM-Newton, an ESA science mission with instruments and contributions directly funded by ESA Member States and NASA.

This research has made use of observations with INTEGRAL, an ESA project with instruments and science data centre funded by ESA member states (especially the PI countries: Denmark, France, Germany, Italy, Switzerland, Spain), and Poland, and with the participation of Russia and the USA.

This research has made use of the NASA/IPAC Extragalactic Database (NED) which is operated by the Jet Propulsion Laboratory, California Institute of Technology, under contract with the National Aeronautics and Space Administration. 

This research has made use of the SIMBAD database, operated at CDS, Strasbourg, France.  

Funding for the SDSS and SDSS-II has been provided by the Alfred P. Sloan Foundation, the Participating Institutions, the National Science Foundation, the U.S. Department of Energy, the National Aeronautics and Space Administration, the Japanese Monbukagakusho, the Max Planck Society, and the Higher Education Funding Council for England. The SDSS Web Site is http://www.sdss.org/.
The SDSS is managed by the Astrophysical Research Consortium for the Participating Institutions. The Participating Institutions are the American Museum of Natural History, Astrophysical Institute Potsdam, University of Basel, University of Cambridge, Case Western Reserve University, University of Chicago, Drexel University, Fermilab, the Institute for Advanced Study, the Japan Participation Group, Johns Hopkins University, the Joint Institute for Nuclear Astrophysics, the Kavli Institute for Particle Astrophysics and Cosmology, the Korean Scientist Group, the Chinese Academy of Sciences (LAMOST), Los Alamos National Laboratory, the Max-Planck-Institute for Astronomy (MPIA), the Max-Planck-Institute for Astrophysics (MPA), New Mexico State University, Ohio State University, University of Pittsburgh, University of Portsmouth, Princeton University, the United States Naval Observatory, and the University of Washington.

Funding for SDSS-III has been provided by the Alfred P. Sloan Foundation, the Participating Institutions, the National Science Foundation, and the U.S. Department of Energy Office of Science. The SDSS-III web site is http://www.sdss3.org/. 
SDSS-III is managed by the Astrophysical Research Consortium for the Participating Institutions of the SDSS-III Collaboration including the University of Arizona, the Brazilian Participation Group, Brookhaven National Laboratory, University of Cambridge, Carnegie Mellon University, University of Florida, the French Participation Group, the German Participation Group, Harvard University, the Instituto de Astrofisica de Canarias, the Michigan State/Notre Dame/JINA Participation Group, Johns Hopkins University, Lawrence Berkeley National Laboratory, Max Planck Institute for Astrophysics, Max Planck Institute for Extraterrestrial Physics, New Mexico State University, New York University, Ohio State University, Pennsylvania State University, University of Portsmouth, Princeton University, the Spanish Participation Group, University of Tokyo, University of Utah, Vanderbilt University, University of Virginia, University of Washington, and Yale University. 

The Digitized Sky Surveys were produced at the Space Telescope Science Institute under U.S. Government grant NAG W-2166. The images of these surveys are based on photographic data obtained using the Oschin Schmidt Telescope on Palomar Mountain and the UK Schmidt Telescope. The plates were processed into the present compressed digital form with the permission of these institutions.

The National Geographic Society - Palomar Observatory Sky Atlas (POSS-I) was made by the California Institute of Technology with grants from the National Geographic Society.

The Second Palomar Observatory Sky Survey (POSS-II) was made by the California Institute of Technology with funds from the National Science Foundation, the National Geographic Society, the Sloan Foundation, the Samuel Oschin Foundation, and the Eastman Kodak Corporation.

The Oschin Schmidt Telescope is operated by the California Institute of Technology and Palomar Observatory.

The UK Schmidt Telescope was operated by the Royal Observatory Edinburgh, with funding from the UK Science and Engineering Research Council (later the UK Particle Physics and Astronomy Research Council), until 1988 June, and thereafter by the Anglo-Australian Observatory. The blue plates of the southern Sky Atlas and its Equatorial Extension (together known as the SERC-J), as well as the Equatorial Red (ER), and the Second Epoch [red] Survey (SES) were all taken with the UK Schmidt.

We acknowledge the efforts of the staff of the Australian Astronomical Observatory (AAO), who developed the 6dF instrument and carried out the observations for the survey.

This publication makes use of data products from the Two Micron All Sky Survey, which is a joint project of the University of Massachusetts and the Infrared Processing and Analysis Center/California Institute of Technology, funded by the National Aeronautics and Space Administration and the National Science Foundation.

This research has made use of NASA's Astrophysics Data System.

\clearpage
\begin{turnpage}
\tabletypesize{\tiny}
\setlength\tabcolsep{0.045in}
\begin{deluxetable}{p{0.4cm}lrrllrrrcrrccccr}

\tablecaption{Catalog of Sources in the 105 Month \textit{Swift}-BAT Survey}
\tablewidth{0pt}
\tablehead{
\colhead{No.\tablenotemark{a}} &
\colhead{BAT Name\tablenotemark{b}} &
\colhead{R.A.\tablenotemark{c}} &
\colhead{Decl.\tablenotemark{c}} &
\colhead{S/N} &
\colhead{Counterpart Name} &
\colhead{R.A.\tablenotemark{c,d}} &
\colhead{Decl.\tablenotemark{c,d}} & 
\colhead{Flux\tablenotemark{e}} &
\colhead{${\rm Flux}_{\rm err}$\tablenotemark{f}} &
\colhead{$\Gamma$\tablenotemark{g}} &
\colhead{$\Gamma_{\rm err}$\tablenotemark{f}} &
\colhead{${\chi_{\rm r}}^2$} &
\colhead{$z$\tablenotemark{h}} &
\colhead{log$L_{\rm BAT}$\tablenotemark{i}} &
\colhead{Cl\tablenotemark{j}} &
\colhead{Type}
}
\startdata
 1&             SWIFT J0001.0-0708&    $0.22826567$&   $-7.16421807$&         7.73&        2MASX J00004876-0709117&        $0.2032$&       $-7.1532$&      13.49&                  $11.10-16.01$&     $2.21$&                    $1.80-2.70$&          1.3&     $0.0375$&        43.64&   50&                Sy1.9\\
 2&             SWIFT J0001.6-7701&     $0.4453023$&  $-77.00031053$&         6.32&                   Fairall 1203&        $0.4419$&    $-76.954002$&      13.23&                  $10.47-15.97$&     $1.62$&                    $1.22-2.04$&          2.2&     $0.0584$&        44.03&   50&                Sy1.9\\
 3&             SWIFT J0002.5+0323&    $0.61264578$&    $3.36496155$&         5.50&                       NGC 7811&        $0.6103$&        $3.3519$&      10.79&                   $7.89-13.64$&     $1.95$&                    $1.38-2.66$&          0.7&     $0.0255$&        43.20&   40&                Sy1.5\\
 4&             SWIFT J0003.3+2737&    $0.85635079$&   $27.64336602$&         5.98&        2MASX J00032742+2739173&        $0.8643$&       $27.6548$&      10.29&                   $7.75-12.77$&     $2.12$&                    $1.61-2.73$&          0.4&     $0.0396$&        43.58&   50&                  Sy2\\
 5&             SWIFT J0005.0+7021&    $0.93403447$&   $70.35778508$&         8.52&        2MASX J00040192+7019185&        $1.0082$&     $70.321701$&      13.41&                  $11.09-15.70$&     $2.07$&                    $1.68-2.52$&          1.0&     $0.0960$&        44.49&   50&                Sy1.9\\
 6&             SWIFT J0006.2+2012&     $1.5962887$&   $20.24152838$&        10.42&                        Mrk 335&        $1.5813$&       $20.2029$&      15.97&                  $13.70-18.44$&     $2.31$&                    $1.97-2.71$&          0.6&     $0.0258$&        43.38&   40&                Sy1.2\\
 7&             SWIFT J0009.4-0037&    $2.30481254$&   $-0.63899431$&         4.26&        2MASX J00091156-0036551&        $2.2982$&       $-0.6152$&      10.17&                   $7.09-13.58$&     $1.68$&                    $1.00-2.43$&          0.3&     $0.0733$&        44.13&   50&                  Sy2\\
 8&             SWIFT J0010.5+1057&    $2.61619405$&   $10.96925199$&        14.02&                       Mrk 1501&        $2.6292$&       $10.9749$&      30.34&                  $27.53-33.40$&     $1.82$&                    $1.61-2.04$&          0.6&     $0.0893$&        44.78&   80&           Beamed AGN\\
 9&             SWIFT J0017.1+8134&    $4.48306744$&   $81.56889298$&         8.57&                [HB89] 0014+813&        $4.2853$&     $81.585602$&      11.39&                   $9.52-13.47$&     $2.42$&                    $1.99-2.95$&          0.7&     $3.3660$&        48.06&   80&           Beamed AGN\\
10&             SWIFT J0021.2-1909&    $5.28922832$&  $-19.16210625$&        10.37&        2MASX J00210753-1910056&        $5.2814$&      $-19.1682$&      19.60&                  $17.51-22.34$&     $2.00$&                    $1.70-2.33$&          0.7&     $0.0956$&        44.65&   50&                  Sy2\\
  $\cdot\cdot\cdot$&   $\cdot\cdot\cdot$&   $\cdot\cdot\cdot$&   $\cdot\cdot\cdot$&   $\cdot\cdot\cdot$&   $\cdot\cdot\cdot$&   $\cdot\cdot\cdot$&   $\cdot\cdot\cdot$&   $\cdot\cdot\cdot$&   $\cdot\cdot\cdot$&   $\cdot\cdot\cdot$&   $\cdot\cdot\cdot$&   $\cdot\cdot\cdot$&   $\cdot\cdot\cdot$&   $\cdot\cdot\cdot$&   $\cdot\cdot\cdot$& $\cdot\cdot\cdot$\\
1211&             SWIFT J0000.5+3251&    $0.12486334$&   $32.84692132$&         5.00&                        IC 5373&      $0.120042$&     $32.782361$&       8.75&                   $6.30-11.38$&     $1.98$&                    $1.39-2.71$&          0.4&     $0.0327$&        43.33&   70&          Unknown AGN\\
1212&             SWIFT J0005.3-7443&      $1.317384$&  $-74.43836876$&         5.69&        2MASX J00052036-7426403&      $1.335083$&      $-74.4445$&      13.38&                  $10.67-16.27$&     $1.47$&                    $1.01-1.93$&          0.5&     $0.1316$&        44.79&   40&                  Sy1\\
1213&             SWIFT J0007.6+0048&    $1.86362056$&    $0.79791497$&         6.14&             SWIFT J0007.6+0048&    $1.86362056$&    $0.79791497$&       4.62&                    $1.72-7.61$&     $2.46$&                    $0.00-0.00$&          1.0&             &             &   15&             multiple\\
1214&             SWIFT J0007.8-4133&    $1.95485005$&  $-41.33825386$&         5.02&             SWIFT J0007.8-4133&    $1.95485005$&  $-41.33825386$&       7.95&                   $5.44-10.59$&     $1.98$&                    $1.20-2.92$&          0.9&             &             &   12&                   U3\\
1215&             SWIFT J0029.1+5937&     $7.2640995$&   $59.37786835$&        13.02&                   V* V1037 Cas&      $7.262833$&        $59.572$&       9.07&                   $6.72-11.55$&     $2.11$&                    $1.59-2.73$&          1.7&             &             &  190&                 LMXB\\
1216&             SWIFT J0043.9-5009&   $10.98664198$&  $-50.15811909$&         4.85&                        NGC 238&     $10.857292$&    $-50.182833$&       5.09&                    $3.43-6.73$&     $3.00$&                    $2.05-4.53$&          0.5&     $0.0287$&        42.98&   70&          Unknown AGN\\
1217&             SWIFT J0046.1-4214&    $11.5214802$&  $-42.14775165$&         5.73&                    PKS 0043-42&     $11.573958$&    $-42.130972$&      10.94&                   $8.30-13.73$&     $1.67$&                    $1.13-2.28$&          0.3&     $0.1160$&        44.58&   60&                LINER\\
1218&             SWIFT J0047.3+1445&    $11.8283861$&    $14.7373579$&         5.18&                      UGC 00488&     $11.830872$&     $14.703535$&      11.28&                   $8.29-13.78$&     $1.67$&                    $1.13-2.27$&          0.6&     $0.0393$&        43.61&   40&                  Sy1\\
1219&             SWIFT J0048.9+8229&   $12.25550056$&   $82.52462277$&         5.90&        2MASX J00502684+8229000&      $12.61275$&     $82.483361$&      10.34&                   $7.95-12.32$&     $1.86$&                    $1.35-2.43$&          0.7&             &             &   70&          Unknown AGN\\
1220&             SWIFT J0052.9+6630&   $13.24044986$&   $66.50208507$&         5.18&        2MASX J00531665+6630336&    $13.3179167$&    $66.5083333$&      10.69&                   $8.06-12.87$&     $1.64$&                    $1.02-2.40$&          1.4&             &             &   11&                   U2\\
  $\cdot\cdot\cdot$&   $\cdot\cdot\cdot$&   $\cdot\cdot\cdot$&   $\cdot\cdot\cdot$&   $\cdot\cdot\cdot$&   $\cdot\cdot\cdot$&   $\cdot\cdot\cdot$&   $\cdot\cdot\cdot$&   $\cdot\cdot\cdot$&   $\cdot\cdot\cdot$&   $\cdot\cdot\cdot$&   $\cdot\cdot\cdot$&   $\cdot\cdot\cdot$&   $\cdot\cdot\cdot$&   $\cdot\cdot\cdot$&   $\cdot\cdot\cdot$& $\cdot\cdot\cdot$
\enddata
\label{tab:catalog}
\tablecomments{(This table is available in its entirety in a machine-readable form in the online journal. A portion is shown here for guidance regarding its form and content.)}
\tablenotetext{a}{BAT number. The provided index is same with that of the 70 month catalog. For the new detections, the index is assigned in the order of increasing right ascension which is given in the third column.}
\tablenotetext{b}{BAT Name.}
\tablenotetext{c}{J2000 coordinates.}
\tablenotetext{d}{Coordinate of counterpart is taken from NED or SIMBAD. If no counterpart is known, the BAT position is listed. In case of source type of U2 whose soft X-ray coordinate is well defined with SNR threshold greater than 3, we provide coordinate of \textit{Swift}-XRT soft X-ray detection ($3-10$ keV).}
\tablenotetext{e}{The flux is extracted from the BAT maps at the position listed for the counterpart, is in units of $10^{-12} {\rm erg s^{-1} cm^{-2}}$, and is computed for the 14-195 keV band.}
\tablenotetext{f}{The error range is the 90\% confidence interval.}
\tablenotetext{g}{The spectral index is computed from a power-law fit to the eight-band BAT data.}
\tablenotetext{h}{The redshifts are taken from the online databases NED and SIMBAD or in a few cases from our own analysis of the optical data. }
\tablenotetext{i}{The luminosity is computed from the flux and redshift in this table, with units of erg ${\rm s^{-1}}$ in the 14-195 keV band.}
\tablenotetext{j}{Source class.}
\tablerefs{\citealt{
Abazajian09, Acero15, Ajello12, Ajello16, Alam15,
Baumgartner13, Beckmann09, Berton15, Bird10, Buttiglione11,
Chavushyan02, Chen15, Coleiro13, Courtois09, Cowperthwaite13, Cusumano10a, Cusumano10b,
DAbrusco14, deRuiter98,
Edelson12, 
Grebenev13,
Haakonsen09, Hakobyan12, Hau95, Healey08, Hiroi13, Huchra12,
Jones09,
Karachentseva10, Karasev12, Kuraszkiewicz09,
Landi17, Loveday96,
Malizia12, Malizia16, Mahony10, Mahony11, Maiorano11, Makarov14, Maselli13, Massaro09, Massaro12, Melendez14, Molina12, Momcheva15,
Oh11, Oh15,
Parisi14, Paturel03,
Raimann05, Ratti10, Rojas17, Roman96, Rousseau00,
Straal16, Sweet14,
Tomsick09, Tomsick12, Tomsick16, Tzanavaris14,
Uzpen07,
Vasudevan13, VeronCetty01, VeronCetty10,
Warwick12, 
Xiong15}}
\end{deluxetable}
\renewcommand{\arraystretch}{1.0}
\clearpage
\end{turnpage}

\bibliographystyle{yahapj}
\bibliography{references}

\begin{thebibliography}{101}
\providecommand\natexlab[1]{#1}
\providecommand\JournalTitle[1]{#1}

\bibitem[{{Abazajian} {et~al.}(2009){Abazajian}, {Adelman-McCarthy},
  {Ag{\"u}eros}, {Allam}, {Allende Prieto}, {An}, {Anderson}, {Anderson},
  {Annis}, {Bahcall}, \& et~al.}]{Abazajian09}
{Abazajian}, K.~N., {Adelman-McCarthy}, J.~K., {Ag{\"u}eros}, M.~A., {et~al.}
  2009,
  \href{http://dx.doi.org/10.1088/0067-0049/182/2/543}{\JournalTitle{\apjs},
  182, 543}

\bibitem[{{Acero} {et~al.}(2015){Acero}, {Ackermann}, {Ajello}, {Albert},
  {Atwood}, {Axelsson}, {Baldini}, {Ballet}, {Barbiellini}, {Bastieri},
  {Belfiore}, {Bellazzini}, {Bissaldi}, {Blandford}, {Bloom}, {Bogart},
  {Bonino}, {Bottacini}, {Bregeon}, {Britto}, {Bruel}, {Buehler}, {Burnett},
  {Buson}, {Caliandro}, {Cameron}, {Caputo}, {Caragiulo}, {Caraveo},
  {Casandjian}, {Cavazzuti}, {Charles}, {Chaves}, {Chekhtman}, {Cheung},
  {Chiang}, {Chiaro}, {Ciprini}, {Claus}, {Cohen-Tanugi}, {Cominsky}, {Conrad},
  {Cutini}, {D'Ammando}, {de Angelis}, {DeKlotz}, {de Palma}, {Desiante},
  {Digel}, {Di Venere}, {Drell}, {Dubois}, {Dumora}, {Favuzzi}, {Fegan},
  {Ferrara}, {Finke}, {Franckowiak}, {Fukazawa}, {Funk}, {Fusco}, {Gargano},
  {Gasparrini}, {Giebels}, {Giglietto}, {Giommi}, {Giordano}, {Giroletti},
  {Glanzman}, {Godfrey}, {Grenier}, {Grondin}, {Grove}, {Guillemot}, {Guiriec},
  {Hadasch}, {Harding}, {Hays}, {Hewitt}, {Hill}, {Horan}, {Iafrate}, {Jogler},
  {J{\'o}hannesson}, {Johnson}, {Johnson}, {Johnson}, {Johnson}, {Kamae},
  {Kataoka}, {Katsuta}, {Kuss}, {La Mura}, {Landriu}, {Larsson}, {Latronico},
  {Lemoine-Goumard}, {Li}, {Li}, {Longo}, {Loparco}, {Lott}, {Lovellette},
  {Lubrano}, {Madejski}, {Massaro}, {Mayer}, {Mazziotta}, {McEnery},
  {Michelson}, {Mirabal}, {Mizuno}, {Moiseev}, {Mongelli}, {Monzani},
  {Morselli}, {Moskalenko}, {Murgia}, {Nuss}, {Ohno}, {Ohsugi}, {Omodei},
  {Orienti}, {Orlando}, {Ormes}, {Paneque}, {Panetta}, {Perkins},
  {Pesce-Rollins}, {Piron}, {Pivato}, {Porter}, {Racusin}, {Rando}, {Razzano},
  {Razzaque}, {Reimer}, {Reimer}, {Reposeur}, {Rochester}, {Romani},
  {Salvetti}, {S{\'a}nchez-Conde}, {Saz Parkinson}, {Schulz}, {Siskind},
  {Smith}, {Spada}, {Spandre}, {Spinelli}, {Stephens}, {Strong}, {Suson},
  {Takahashi}, {Takahashi}, {Tanaka}, {Thayer}, {Thayer}, {Thompson},
  {Tibaldo}, {Tibolla}, {Torres}, {Torresi}, {Tosti}, {Troja}, {Van Klaveren},
  {Vianello}, {Winer}, {Wood}, {Wood}, {Zimmer}, \& {Fermi-LAT
  Collaboration}}]{Acero15}
{Acero}, F., {Ackermann}, M., {Ajello}, M., {et~al.} 2015,
  \href{http://dx.doi.org/10.1088/0067-0049/218/2/23}{\JournalTitle{\apjs},
  218, 23}

\bibitem[{{Ajello} {et~al.}(2012){Ajello}, {Alexander}, {Greiner}, {Madejski},
  {Gehrels}, \& {Burlon}}]{Ajello12}
{Ajello}, M., {Alexander}, D.~M., {Greiner}, J., {et~al.} 2012,
  \href{http://dx.doi.org/10.1088/0004-637X/749/1/21}{\JournalTitle{\apj}, 749,
  21}

\bibitem[{{Ajello} {et~al.}(2016){Ajello}, {Ghisellini}, {Paliya}, {Kocevski},
  {Tagliaferri}, {Madejski}, {Rau}, {Schady}, {Greiner}, {Massaro},
  {Balokovi{\'c}}, {B{\"u}hler}, {Giomi}, {Marcotulli}, {D'Ammando}, {Stern},
  {Boggs}, {Christensen}, {Craig}, {Hailey}, {Harrison}, \& {Zhang}}]{Ajello16}
{Ajello}, M., {Ghisellini}, G., {Paliya}, V.~S., {et~al.} 2016,
  \href{http://dx.doi.org/10.3847/0004-637X/826/1/76}{\JournalTitle{\apj}, 826,
  76}

\bibitem[{{Alam} {et~al.}(2015){Alam}, {Albareti}, {Allende Prieto}, {Anders},
  {Anderson}, {Anderton}, {Andrews}, {Armengaud}, {Aubourg}, {Bailey}, \&
  et~al.}]{Alam15}
{Alam}, S., {Albareti}, F.~D., {Allende Prieto}, C., {et~al.} 2015,
  \href{http://dx.doi.org/10.1088/0067-0049/219/1/12}{\JournalTitle{\apjs},
  219, 12}

\bibitem[{{Arnaud}(1996)}]{Arnaud96}
{Arnaud}, K.~A. 1996, in Astronomical Society of the Pacific Conference Series,
  Vol. 101, Astronomical Data Analysis Software and Systems V, ed. G.~H.
  {Jacoby} \& J.~{Barnes}, 17

\bibitem[{{Barthelmy} {et~al.}(2005){Barthelmy}, {Barbier}, {Cummings},
  {Fenimore}, {Gehrels}, {Hullinger}, {Krimm}, {Markwardt}, {Palmer},
  {Parsons}, {Sato}, {Suzuki}, {Takahashi}, {Tashiro}, \&
  {Tueller}}]{Barthelmy05}
{Barthelmy}, S.~D., {Barbier}, L.~M., {Cummings}, J.~R., {et~al.} 2005,
  \href{http://dx.doi.org/10.1007/s11214-005-5096-3}{\JournalTitle{\ssr}, 120,
  143}

\bibitem[{{Baumgartner} {et~al.}(2013){Baumgartner}, {Tueller}, {Markwardt},
  {Skinner}, {Barthelmy}, {Mushotzky}, {Evans}, \& {Gehrels}}]{Baumgartner13}
{Baumgartner}, W.~H., {Tueller}, J., {Markwardt}, C.~B., {et~al.} 2013,
  \href{http://dx.doi.org/10.1088/0067-0049/207/2/19}{\JournalTitle{\apjs},
  207, 19}

\bibitem[{{Beckmann} {et~al.}(2009){Beckmann}, {Soldi}, {Ricci},
  {Alfonso-Garz{\'o}n}, {Courvoisier}, {Domingo}, {Gehrels}, {Lubi{\'n}ski},
  {Mas-Hesse}, \& {Zdziarski}}]{Beckmann09}
{Beckmann}, V., {Soldi}, S., {Ricci}, C., {et~al.} 2009,
  \href{http://dx.doi.org/10.1051/0004-6361/200912111}{\JournalTitle{\aap},
  505, 417}

\bibitem[{{Berney} {et~al.}(2015){Berney}, {Koss}, {Trakhtenbrot}, {Ricci},
  {Lamperti}, {Schawinski}, {Balokovi{\'c}}, {Crenshaw}, {Fischer}, {Gehrels},
  {Harrison}, {Hashimoto}, {Ichikawa}, {Mushotzky}, {Oh}, {Stern}, {Treister},
  {Ueda}, {Veilleux}, \& {Winter}}]{Berney15}
{Berney}, S., {Koss}, M., {Trakhtenbrot}, B., {et~al.} 2015,
  \href{http://dx.doi.org/10.1093/mnras/stv2181}{\JournalTitle{\mnras}, 454,
  3622}

\bibitem[{{Berton} {et~al.}(2015){Berton}, {Foschini}, {Ciroi}, {Cracco}, {La
  Mura}, {Lister}, {Mathur}, {Peterson}, {Richards}, \& {Rafanelli}}]{Berton15}
{Berton}, M., {Foschini}, L., {Ciroi}, S., {et~al.} 2015,
  \href{http://dx.doi.org/10.1051/0004-6361/201525691}{\JournalTitle{\aap},
  578, A28}

\bibitem[{{Bird} {et~al.}(2010){Bird}, {Bazzano}, {Bassani}, {Capitanio},
  {Fiocchi}, {Hill}, {Malizia}, {McBride}, {Scaringi}, {Sguera}, {Stephen},
  {Ubertini}, {Dean}, {Lebrun}, {Terrier}, {Renaud}, {Mattana}, {G{\"o}tz},
  {Rodriguez}, {Belanger}, {Walter}, \& {Winkler}}]{Bird10}
{Bird}, A.~J., {Bazzano}, A., {Bassani}, L., {et~al.} 2010,
  \href{http://dx.doi.org/10.1088/0067-0049/186/1/1}{\JournalTitle{\apjs}, 186,
  1}

\bibitem[{{Bird} {et~al.}(2016){Bird}, {Bazzano}, {Malizia}, {Fiocchi},
  {Sguera}, {Bassani}, {Hill}, {Ubertini}, \& {Winkler}}]{Bird16}
{Bird}, A.~J., {Bazzano}, A., {Malizia}, A., {et~al.} 2016,
  \href{http://dx.doi.org/10.3847/0067-0049/223/1/15}{\JournalTitle{\apjs},
  223, 15}

\bibitem[{{Boller} {et~al.}(2016){Boller}, {Freyberg}, {Tr{\"u}mper}, {Haberl},
  {Voges}, \& {Nandra}}]{Boller16}
{Boller}, T., {Freyberg}, M.~J., {Tr{\"u}mper}, J., {et~al.} 2016,
  \href{http://dx.doi.org/10.1051/0004-6361/201525648}{\JournalTitle{\aap},
  588, A103}

\bibitem[{{Burrows} {et~al.}(2005){Burrows}, {Hill}, {Nousek}, {Kennea},
  {Wells}, {Osborne}, {Abbey}, {Beardmore}, {Mukerjee}, {Short}, {Chincarini},
  {Campana}, {Citterio}, {Moretti}, {Pagani}, {Tagliaferri}, {Giommi},
  {Capalbi}, {Tamburelli}, {Angelini}, {Cusumano}, {Br{\"a}uninger}, {Burkert},
  \& {Hartner}}]{Burrows05}
{Burrows}, D.~N., {Hill}, J.~E., {Nousek}, J.~A., {et~al.} 2005,
  \href{http://dx.doi.org/10.1007/s11214-005-5097-2}{\JournalTitle{\ssr}, 120,
  165}

\bibitem[{{Buttiglione} {et~al.}(2011){Buttiglione}, {Capetti}, {Celotti},
  {Axon}, {Chiaberge}, {Macchetto}, \& {Sparks}}]{Buttiglione11}
{Buttiglione}, S., {Capetti}, A., {Celotti}, A., {et~al.} 2011,
  \href{http://dx.doi.org/10.1051/0004-6361/201015574}{\JournalTitle{\aap},
  525, A28}

\bibitem[{{Chambers} {et~al.}(2016){Chambers}, {Magnier}, {Metcalfe},
  {Flewelling}, {Huber}, {Waters}, {Denneau}, {Draper}, {Farrow}, {Finkbeiner},
  {Holmberg}, {Koppenhoefer}, {Price}, {Saglia}, {Schlafly}, {Smartt},
  {Sweeney}, {Wainscoat}, {Burgett}, {Grav}, {Heasley}, {Hodapp}, {Jedicke},
  {Kaiser}, {Kudritzki}, {Luppino}, {Lupton}, {Monet}, {Morgan}, {Onaka},
  {Stubbs}, {Tonry}, {Banados}, {Bell}, {Bender}, {Bernard}, {Botticella},
  {Casertano}, {Chastel}, {Chen}, {Chen}, {Cole}, {Deacon}, {Frenk},
  {Fitzsimmons}, {Gezari}, {Goessl}, {Goggia}, {Goldman}, {Grebel}, {Hambly},
  {Hasinger}, {Heavens}, {Heckman}, {Henderson}, {Henning}, {Holman}, {Hopp},
  {Ip}, {Isani}, {Keyes}, {Koekemoer}, {Kotak}, {Long}, {Lucey}, {Liu},
  {Martin}, {McLean}, {Morganson}, {Murphy}, {Nieto-Santisteban}, {Norberg},
  {Peacock}, {Pier}, {Postman}, {Primak}, {Rae}, {Rest}, {Riess}, {Riffeser},
  {Rix}, {Roser}, {Schilbach}, {Schultz}, {Scolnic}, {Szalay}, {Seitz},
  {Shiao}, {Small}, {Smith}, {Soderblom}, {Taylor}, {Thakar}, {Thiel},
  {Thilker}, {Urata}, {Valenti}, {Walter}, {Watters}, {Werner}, {White},
  {Wood-Vasey}, \& {Wyse}}]{Chambers16}
{Chambers}, K.~C., {Magnier}, E.~A., {Metcalfe}, N., {et~al.} 2016,
  \JournalTitle{ArXiv e-prints},
  \href{http://arxiv.org/abs/1612.05560}{{\sffamily arXiv:1612.05560
  [astro-ph.IM]}}

\bibitem[{{Chavushyan} {et~al.}(2002){Chavushyan}, {Mujica}, {Valdes},
  {Gorshkov}, {Konnikova}, \& {Mingaliev}}]{Chavushyan02}
{Chavushyan}, V., {Mujica}, R., {Valdes}, J.~R., {et~al.} 2002,
  \href{http://dx.doi.org/10.1134/1.1508061}{\JournalTitle{Astronomy Reports},
  46, 697}

\bibitem[{{Chen} {et~al.}(2015){Chen}, {Gu}, \& {Chen}}]{Chen15}
{Chen}, Z.-F., {Gu}, Q.-S., \& {Chen}, Y.-M. 2015,
  \href{http://dx.doi.org/10.1088/0067-0049/221/2/32}{\JournalTitle{\apjs},
  221, 32}

\bibitem[{{Coleiro} {et~al.}(2013){Coleiro}, {Chaty}, {Zurita Heras}, {Rahoui},
  \& {Tomsick}}]{Coleiro13}
{Coleiro}, A., {Chaty}, S., {Zurita Heras}, J.~A., {Rahoui}, F., \& {Tomsick},
  J.~A. 2013,
  \href{http://dx.doi.org/10.1051/0004-6361/201322382}{\JournalTitle{\aap},
  560, A108}

\bibitem[{{Courtois} {et~al.}(2009){Courtois}, {Tully}, {Fisher}, {Bonhomme},
  {Zavodny}, \& {Barnes}}]{Courtois09}
{Courtois}, H.~M., {Tully}, R.~B., {Fisher}, J.~R., {et~al.} 2009,
  \href{http://dx.doi.org/10.1088/0004-6256/138/6/1938}{\JournalTitle{\aj},
  138, 1938}

\bibitem[{{Cowperthwaite} {et~al.}(2013){Cowperthwaite}, {Massaro},
  {D'Abrusco}, {Paggi}, {Tosti}, \& {Smith}}]{Cowperthwaite13}
{Cowperthwaite}, P.~S., {Massaro}, F., {D'Abrusco}, R., {et~al.} 2013,
  \href{http://dx.doi.org/10.1088/0004-6256/146/5/110}{\JournalTitle{\aj}, 146,
  110}

\bibitem[{{Cusumano} {et~al.}(2010{\natexlab{a}}){Cusumano}, {La Parola},
  {Segreto}, {Mangano}, {Ferrigno}, {Maselli}, {Romano}, {Mineo}, {Sbarufatti},
  {Campana}, {Chincarini}, {Giommi}, {Masetti}, {Moretti}, \&
  {Tagliaferri}}]{Cusumano10a}
{Cusumano}, G., {La Parola}, V., {Segreto}, A., {et~al.} 2010{\natexlab{a}},
  \href{http://dx.doi.org/10.1051/0004-6361/200811184}{\JournalTitle{\aap},
  510, A48}

\bibitem[{{Cusumano} {et~al.}(2010{\natexlab{b}}){Cusumano}, {La Parola},
  {Segreto}, {Ferrigno}, {Maselli}, {Sbarufatti}, {Romano}, {Chincarini},
  {Giommi}, {Masetti}, {Moretti}, {Parisi}, \& {Tagliaferri}}]{Cusumano10b}
---. 2010{\natexlab{b}},
  \href{http://dx.doi.org/10.1051/0004-6361/201015249}{\JournalTitle{\aap},
  524, A64}

\bibitem[{{D'Abrusco} {et~al.}(2014){D'Abrusco}, {Massaro}, {Paggi}, {Smith},
  {Masetti}, {Landoni}, \& {Tosti}}]{DAbrusco14}
{D'Abrusco}, R., {Massaro}, F., {Paggi}, A., {et~al.} 2014,
  \href{http://dx.doi.org/10.1088/0067-0049/215/1/14}{\JournalTitle{\apjs},
  215, 14}

\bibitem[{{de Ruiter} {et~al.}(1998){de Ruiter}, {Parma}, {Stirpe},
  {Perez-Fournon}, {Gonzalez-Serrano}, {Rengelink}, \& {Bremer}}]{deRuiter98}
{de Ruiter}, H.~R., {Parma}, P., {Stirpe}, G.~M., {et~al.} 1998,
  \JournalTitle{\aap}, 339, 34

\bibitem[{{Edelson} \& {Malkan}(2012)}]{Edelson12}
{Edelson}, R., \& {Malkan}, M. 2012,
  \href{http://dx.doi.org/10.1088/0004-637X/751/1/52}{\JournalTitle{\apj}, 751,
  52}

\bibitem[{{Forman} {et~al.}(1978){Forman}, {Jones}, {Cominsky}, {Julien},
  {Murray}, {Peters}, {Tananbaum}, \& {Giacconi}}]{Forman78}
{Forman}, W., {Jones}, C., {Cominsky}, L., {et~al.} 1978,
  \href{http://dx.doi.org/10.1086/190561}{\JournalTitle{\apjs}, 38, 357}

\bibitem[{{Gehrels} {et~al.}(2004){Gehrels}, {Chincarini}, {Giommi}, {Mason},
  {Nousek}, {Wells}, {White}, {Barthelmy}, {Burrows}, {Cominsky}, {Hurley},
  {Marshall}, {M{\'e}sz{\'a}ros}, {Roming}, {Angelini}, {Barbier}, {Belloni},
  {Campana}, {Caraveo}, {Chester}, {Citterio}, {Cline}, {Cropper}, {Cummings},
  {Dean}, {Feigelson}, {Fenimore}, {Frail}, {Fruchter}, {Garmire}, {Gendreau},
  {Ghisellini}, {Greiner}, {Hill}, {Hunsberger}, {Krimm}, {Kulkarni}, {Kumar},
  {Lebrun}, {Lloyd-Ronning}, {Markwardt}, {Mattson}, {Mushotzky}, {Norris},
  {Osborne}, {Paczynski}, {Palmer}, {Park}, {Parsons}, {Paul}, {Rees},
  {Reynolds}, {Rhoads}, {Sasseen}, {Schaefer}, {Short}, {Smale}, {Smith},
  {Stella}, {Tagliaferri}, {Takahashi}, {Tashiro}, {Townsley}, {Tueller},
  {Turner}, {Vietri}, {Voges}, {Ward}, {Willingale}, {Zerbi}, \&
  {Zhang}}]{Gehrels04}
{Gehrels}, N., {Chincarini}, G., {Giommi}, P., {et~al.} 2004,
  \href{http://dx.doi.org/10.1086/422091}{\JournalTitle{\apj}, 611, 1005}

\bibitem[{{Giacconi} {et~al.}(1971){Giacconi}, {Kellogg}, {Gorenstein},
  {Gursky}, \& {Tananbaum}}]{Giacconi71}
{Giacconi}, R., {Kellogg}, E., {Gorenstein}, P., {Gursky}, H., \& {Tananbaum},
  H. 1971, \href{http://dx.doi.org/10.1086/180711}{\JournalTitle{\apjl}, 165,
  L27}

\bibitem[{{Grebenev} {et~al.}(2013){Grebenev}, {Lutovinov}, {Tsygankov}, \&
  {Mereminskiy}}]{Grebenev13}
{Grebenev}, S.~A., {Lutovinov}, A.~A., {Tsygankov}, S.~S., \& {Mereminskiy},
  I.~A. 2013,
  \href{http://dx.doi.org/10.1093/mnras/sts008}{\JournalTitle{\mnras}, 428, 50}

\bibitem[{{Haakonsen} \& {Rutledge}(2009)}]{Haakonsen09}
{Haakonsen}, C.~B., \& {Rutledge}, R.~E. 2009,
  \href{http://dx.doi.org/10.1088/0067-0049/184/1/138}{\JournalTitle{\apjs},
  184, 138}

\bibitem[{{Hakobyan} {et~al.}(2012){Hakobyan}, {Adibekyan}, {Aramyan},
  {Petrosian}, {Gomes}, {Mamon}, {Kunth}, \& {Turatto}}]{Hakobyan12}
{Hakobyan}, A.~A., {Adibekyan}, V.~Z., {Aramyan}, L.~S., {et~al.} 2012,
  \href{http://dx.doi.org/10.1051/0004-6361/201219541}{\JournalTitle{\aap},
  544, A81}

\bibitem[{{Hau} {et~al.}(1995){Hau}, {Ferguson}, {Lahav}, \&
  {Lynden-Bell}}]{Hau95}
{Hau}, G.~K.~T., {Ferguson}, H.~C., {Lahav}, O., \& {Lynden-Bell}, D. 1995,
  \href{http://dx.doi.org/10.1093/mnras/277.1.125}{\JournalTitle{\mnras}, 277,
  125}

\bibitem[{{Healey} {et~al.}(2008){Healey}, {Romani}, {Cotter}, {Michelson},
  {Schlafly}, {Readhead}, {Giommi}, {Chaty}, {Grenier}, \&
  {Weintraub}}]{Healey08}
{Healey}, S.~E., {Romani}, R.~W., {Cotter}, G., {et~al.} 2008,
  \href{http://dx.doi.org/10.1086/523302}{\JournalTitle{\apjs}, 175, 97}

\bibitem[{{Hiroi} {et~al.}(2013){Hiroi}, {Ueda}, {Hayashida}, {Shidatsu},
  {Sato}, {Kawamuro}, {Sugizaki}, {Nakahira}, {Serino}, {Kawai}, {Matsuoka},
  {Mihara}, {Morii}, {Nakajima}, {Negoro}, {Sakamoto}, {Tomida}, {Tsuboi},
  {Tsunemi}, {Ueno}, {Yamaoka}, {Yoshida}, {Asada}, {Eguchi}, {Hanayama},
  {Higa}, {Ishikawa}, {Ishikawa}, {Isobe}, {Kohama}, {Kimura}, {Morihana},
  {Nakagawa}, {Nakano}, {Nishimura}, {Ogawa}, {Sasaki}, {Sugimoto}, {Takagi},
  {Usui}, {Yamamoto}, {Yamauchi}, \& {Yoshidome}}]{Hiroi13}
{Hiroi}, K., {Ueda}, Y., {Hayashida}, M., {et~al.} 2013,
  \href{http://dx.doi.org/10.1088/0067-0049/207/2/36}{\JournalTitle{\apjs},
  207, 36}

\bibitem[{{Huchra} {et~al.}(2012){Huchra}, {Macri}, {Masters}, {Jarrett},
  {Berlind}, {Calkins}, {Crook}, {Cutri}, {Erdo{\v g}du}, {Falco}, {George},
  {Hutcheson}, {Lahav}, {Mader}, {Mink}, {Martimbeau}, {Schneider},
  {Skrutskie}, {Tokarz}, \& {Westover}}]{Huchra12}
{Huchra}, J.~P., {Macri}, L.~M., {Masters}, K.~L., {et~al.} 2012,
  \href{http://dx.doi.org/10.1088/0067-0049/199/2/26}{\JournalTitle{\apjs},
  199, 26}

\bibitem[{{Jones} {et~al.}(2009){Jones}, {Read}, {Saunders}, {Colless},
  {Jarrett}, {Parker}, {Fairall}, {Mauch}, {Sadler}, {Watson}, {Burton},
  {Campbell}, {Cass}, {Croom}, {Dawe}, {Fiegert}, {Frankcombe}, {Hartley},
  {Huchra}, {James}, {Kirby}, {Lahav}, {Lucey}, {Mamon}, {Moore}, {Peterson},
  {Prior}, {Proust}, {Russell}, {Safouris}, {Wakamatsu}, {Westra}, \&
  {Williams}}]{Jones09}
{Jones}, D.~H., {Read}, M.~A., {Saunders}, W., {et~al.} 2009,
  \href{http://dx.doi.org/10.1111/j.1365-2966.2009.15338.x}{\JournalTitle{\mnras},
  399, 683}

\bibitem[{{Karachentseva} {et~al.}(2010){Karachentseva}, {Mitronova}, {Melnyk},
  \& {Karachentsev}}]{Karachentseva10}
{Karachentseva}, V.~E., {Mitronova}, S.~N., {Melnyk}, O.~V., \& {Karachentsev},
  I.~D. 2010,
  \href{http://dx.doi.org/10.1134/S1990341310010013}{\JournalTitle{Astrophysical
  Bulletin}, 65, 1}

\bibitem[{{Karasev} {et~al.}(2012){Karasev}, {Lutovinov}, {Revnivtsev}, \&
  {Krivonos}}]{Karasev12}
{Karasev}, D.~I., {Lutovinov}, A.~A., {Revnivtsev}, M.~G., \& {Krivonos}, R.~A.
  2012,
  \href{http://dx.doi.org/10.1134/S1063773712100039}{\JournalTitle{Astronomy
  Letters}, 38, 629}

\bibitem[{{Koss} {et~al.}(2013){Koss}, {Mushotzky}, {Baumgartner}, {Veilleux},
  {Tueller}, {Markwardt}, \& {Casey}}]{Koss13}
{Koss}, M., {Mushotzky}, R., {Baumgartner}, W., {et~al.} 2013,
  \href{http://dx.doi.org/10.1088/2041-8205/765/2/L26}{\JournalTitle{\apjl},
  765, L26}

\bibitem[{{Koss} {et~al.}(2012){Koss}, {Mushotzky}, {Treister}, {Veilleux},
  {Vasudevan}, \& {Trippe}}]{Koss12}
{Koss}, M., {Mushotzky}, R., {Treister}, E., {et~al.} 2012,
  \href{http://dx.doi.org/10.1088/2041-8205/746/2/L22}{\JournalTitle{\apjl},
  746, L22}

\bibitem[{{Koss} {et~al.}(2017){Koss}, {Trakhtenbrot}, {Ricci}, {Lamperti},
  {Oh}, {Berney}, {Schawinski}, {Balokovic}, {Baronchelli}, {Crenshaw},
  {Fischer}, {Gehrels}, {Harrison}, {Hashimoto}, {Hogg}, {Ichikawa}, {Masetti},
  {Mushotzky}, {Stern}, {Treister}, {Ueda}, {Veilleux}, \& {Winter}}]{Koss17}
{Koss}, M., {Trakhtenbrot}, B., {Ricci}, C., {et~al.} 2017, \JournalTitle{ArXiv
  e-prints}, \href{http://arxiv.org/abs/1707.08123}{{\sffamily arXiv:1707.08123
  [astro-ph.HE]}}

\bibitem[{{Koss} {et~al.}(2016){Koss}, {Assef}, {Balokovi{\'c}}, {Stern},
  {Gandhi}, {Lamperti}, {Alexander}, {Ballantyne}, {Bauer}, {Berney}, {Brandt},
  {Comastri}, {Gehrels}, {Harrison}, {Lansbury}, {Markwardt}, {Ricci},
  {Rivers}, {Schawinski}, {Trakhtenbrot}, {Treister}, \& {Urry}}]{Koss16}
{Koss}, M.~J., {Assef}, R., {Balokovi{\'c}}, M., {et~al.} 2016,
  \href{http://dx.doi.org/10.3847/0004-637X/825/2/85}{\JournalTitle{\apj}, 825,
  85}

\bibitem[{{Krivonos} {et~al.}(2010){Krivonos}, {Tsygankov}, {Revnivtsev},
  {Grebenev}, {Churazov}, \& {Sunyaev}}]{Krivonos10}
{Krivonos}, R., {Tsygankov}, S., {Revnivtsev}, M., {et~al.} 2010,
  \href{http://dx.doi.org/10.1051/0004-6361/201014935}{\JournalTitle{\aap},
  523, A61}

\bibitem[{{Kuraszkiewicz} {et~al.}(2009){Kuraszkiewicz}, {Wilkes}, {Schmidt},
  {Ghosh}, {Smith}, {Cutri}, {Hines}, {Huff}, {McDowell}, \&
  {Nelson}}]{Kuraszkiewicz09}
{Kuraszkiewicz}, J., {Wilkes}, B.~J., {Schmidt}, G., {et~al.} 2009,
  \href{http://dx.doi.org/10.1088/0004-637X/692/2/1143}{\JournalTitle{\apj},
  692, 1143}

\bibitem[{{Lamperti} {et~al.}(2017){Lamperti}, {Koss}, {Trakhtenbrot},
  {Schawinski}, {Ricci}, {Oh}, {Landt}, {Riffel}, {Rodr{\'{\i}}guez-Ardila},
  {Gehrels}, {Harrison}, {Masetti}, {Mushotzky}, {Treister}, {Ueda}, \&
  {Veilleux}}]{Lamperti17}
{Lamperti}, I., {Koss}, M., {Trakhtenbrot}, B., {et~al.} 2017,
  \href{http://dx.doi.org/10.1093/mnras/stx055}{\JournalTitle{\mnras}, 467,
  540}

\bibitem[{{Landi} {et~al.}(2017){Landi}, {Bassani}, {Bazzano}, {Bird},
  {Fiocchi}, {Malizia}, {Panessa}, {Sguera}, \& {Ubertini}}]{Landi17}
{Landi}, R., {Bassani}, L., {Bazzano}, A., {et~al.} 2017, \JournalTitle{ArXiv
  e-prints}, \href{http://arxiv.org/abs/1704.03872}{{\sffamily arXiv:1704.03872
  [astro-ph.HE]}}

\bibitem[{{Levine} {et~al.}(1984){Levine}, {Lang}, {Lewin}, {Primini},
  {Dobson}, {Doty}, {Hoffman}, {Howe}, {Scheepmaker}, {Wheaton}, {Matteson},
  {Baity}, {Gruber}, {Knight}, {Nolan}, {Pelling}, {Rothschild}, \&
  {Peterson}}]{Levine84}
{Levine}, A.~M., {Lang}, F.~L., {Lewin}, W.~H.~G., {et~al.} 1984,
  \href{http://dx.doi.org/10.1086/190944}{\JournalTitle{\apjs}, 54, 581}

\bibitem[{{Lien} {et~al.}(2016){Lien}, {Sakamoto}, {Barthelmy}, {Baumgartner},
  {Cannizzo}, {Chen}, {Collins}, {Cummings}, {Gehrels}, {Krimm}, {Markwardt},
  {Palmer}, {Stamatikos}, {Troja}, \& {Ukwatta}}]{Lien16}
{Lien}, A., {Sakamoto}, T., {Barthelmy}, S.~D., {et~al.} 2016,
  \href{http://dx.doi.org/10.3847/0004-637X/829/1/7}{\JournalTitle{\apj}, 829,
  7}

\bibitem[{{Loveday}(1996)}]{Loveday96}
{Loveday}, J. 1996,
  \href{http://dx.doi.org/10.1093/mnras/278.4.1025}{\JournalTitle{\mnras}, 278,
  1025}

\bibitem[{{Mahony} {et~al.}(2010){Mahony}, {Croom}, {Boyle}, {Edge}, {Mauch},
  \& {Sadler}}]{Mahony10}
{Mahony}, E.~K., {Croom}, S.~M., {Boyle}, B.~J., {et~al.} 2010,
  \href{http://dx.doi.org/10.1111/j.1365-2966.2009.15705.x}{\JournalTitle{\mnras},
  401, 1151}

\bibitem[{{Mahony} {et~al.}(2011){Mahony}, {Sadler}, {Croom}, {Ekers},
  {Bannister}, {Chhetri}, {Hancock}, {Johnston}, {Massardi}, \&
  {Murphy}}]{Mahony11}
{Mahony}, E.~K., {Sadler}, E.~M., {Croom}, S.~M., {et~al.} 2011,
  \href{http://dx.doi.org/10.1111/j.1365-2966.2011.19427.x}{\JournalTitle{\mnras},
  417, 2651}

\bibitem[{{Maiorano} {et~al.}(2011){Maiorano}, {Landi}, {Stephen}, {Bassani},
  {Masetti}, {Parisi}, {Palazzi}, {Parma}, {Bird}, {Bazzano}, {Ubertini},
  {Jim{\'e}nez-Bail{\'o}n}, {Chavushyan}, {Galaz}, {Minniti}, \&
  {Morelli}}]{Maiorano11}
{Maiorano}, E., {Landi}, R., {Stephen}, J.~B., {et~al.} 2011,
  \href{http://dx.doi.org/10.1111/j.1365-2966.2011.19065.x}{\JournalTitle{\mnras},
  416, 531}

\bibitem[{{Makarov} {et~al.}(2014){Makarov}, {Prugniel}, {Terekhova},
  {Courtois}, \& {Vauglin}}]{Makarov14}
{Makarov}, D., {Prugniel}, P., {Terekhova}, N., {Courtois}, H., \& {Vauglin},
  I. 2014,
  \href{http://dx.doi.org/10.1051/0004-6361/201423496}{\JournalTitle{\aap},
  570, A13}

\bibitem[{{Malizia} {et~al.}(2012){Malizia}, {Bassani}, {Bazzano}, {Bird},
  {Masetti}, {Panessa}, {Stephen}, \& {Ubertini}}]{Malizia12}
{Malizia}, A., {Bassani}, L., {Bazzano}, A., {et~al.} 2012,
  \href{http://dx.doi.org/10.1111/j.1365-2966.2012.21755.x}{\JournalTitle{\mnras},
  426, 1750}

\bibitem[{{Malizia} {et~al.}(2016){Malizia}, {Landi}, {Molina}, {Bassani},
  {Bazzano}, {Bird}, \& {Ubertini}}]{Malizia16}
{Malizia}, A., {Landi}, R., {Molina}, M., {et~al.} 2016,
  \href{http://dx.doi.org/10.1093/mnras/stw972}{\JournalTitle{\mnras}, 460, 19}

\bibitem[{{Markwardt} {et~al.}(2005){Markwardt}, {Tueller}, {Skinner},
  {Gehrels}, {Barthelmy}, \& {Mushotzky}}]{Markwardt05}
{Markwardt}, C.~B., {Tueller}, J., {Skinner}, G.~K., {et~al.} 2005,
  \href{http://dx.doi.org/10.1086/498569}{\JournalTitle{\apjl}, 633, L77}

\bibitem[{{Maselli} {et~al.}(2013){Maselli}, {Massaro}, {Cusumano},
  {D'Abrusco}, {La Parola}, {Paggi}, {Segreto}, {Smith}, \&
  {Tosti}}]{Maselli13}
{Maselli}, A., {Massaro}, F., {Cusumano}, G., {et~al.} 2013,
  \href{http://dx.doi.org/10.1088/0067-0049/206/2/17}{\JournalTitle{\apjs},
  206, 17}

\bibitem[{{Massaro} {et~al.}(2009){Massaro}, {Giommi}, {Leto}, {Marchegiani},
  {Maselli}, {Perri}, {Piranomonte}, \& {Sclavi}}]{Massaro09}
{Massaro}, E., {Giommi}, P., {Leto}, C., {et~al.} 2009,
  \href{http://dx.doi.org/10.1051/0004-6361:200810161}{\JournalTitle{\aap},
  495, 691}

\bibitem[{{Massaro} {et~al.}(2012){Massaro}, {Paggi}, {D'Abrusco}, \&
  {Tosti}}]{Massaro12}
{Massaro}, F., {Paggi}, A., {D'Abrusco}, R., \& {Tosti}, G. 2012,
  \href{http://dx.doi.org/10.1088/2041-8205/750/2/L35}{\JournalTitle{\apjl},
  750, L35}

\bibitem[{{Meegan} {et~al.}(2009){Meegan}, {Lichti}, {Bhat}, {Bissaldi},
  {Briggs}, {Connaughton}, {Diehl}, {Fishman}, {Greiner}, {Hoover}, {van der
  Horst}, {von Kienlin}, {Kippen}, {Kouveliotou}, {McBreen}, {Paciesas},
  {Preece}, {Steinle}, {Wallace}, {Wilson}, \& {Wilson-Hodge}}]{Meegan09}
{Meegan}, C., {Lichti}, G., {Bhat}, P.~N., {et~al.} 2009,
  \href{http://dx.doi.org/10.1088/0004-637X/702/1/791}{\JournalTitle{\apj},
  702, 791}

\bibitem[{{Mel{\'e}ndez} {et~al.}(2014){Mel{\'e}ndez}, {Mushotzky}, {Shimizu},
  {Barger}, \& {Cowie}}]{Melendez14}
{Mel{\'e}ndez}, M., {Mushotzky}, R.~F., {Shimizu}, T.~T., {Barger}, A.~J., \&
  {Cowie}, L.~L. 2014,
  \href{http://dx.doi.org/10.1088/0004-637X/794/2/152}{\JournalTitle{\apj},
  794, 152}

\bibitem[{{Molina} {et~al.}(2012){Molina}, {Landi}, {Bassani}, {Bazzano},
  {Fiocchi}, {Bird}, \& {Drave}}]{Molina12}
{Molina}, M., {Landi}, R., {Bassani}, L., {et~al.} 2012, \JournalTitle{The
  Astronomer's Telegram}, 4250

\bibitem[{{Momcheva} {et~al.}(2015){Momcheva}, {Williams}, {Cool}, {Keeton}, \&
  {Zabludoff}}]{Momcheva15}
{Momcheva}, I.~G., {Williams}, K.~A., {Cool}, R.~J., {Keeton}, C.~R., \&
  {Zabludoff}, A.~I. 2015,
  \href{http://dx.doi.org/10.1088/0067-0049/219/2/29}{\JournalTitle{\apjs},
  219, 29}

\bibitem[{{Oh} {et~al.}(2011){Oh}, {Sarzi}, {Schawinski}, \& {Yi}}]{Oh11}
{Oh}, K., {Sarzi}, M., {Schawinski}, K., \& {Yi}, S.~K. 2011,
  \href{http://dx.doi.org/10.1088/0067-0049/195/2/13}{\JournalTitle{\apjs},
  195, 13}

\bibitem[{{Oh} {et~al.}(2015){Oh}, {Yi}, {Schawinski}, {Koss}, {Trakhtenbrot},
  \& {Soto}}]{Oh15}
{Oh}, K., {Yi}, S.~K., {Schawinski}, K., {et~al.} 2015,
  \href{http://dx.doi.org/10.1088/0067-0049/219/1/1}{\JournalTitle{\apjs}, 219,
  1}

\bibitem[{{Oh} {et~al.}(2017){Oh}, {Schawinski}, {Koss}, {Trakhtenbrot},
  {Lamperti}, {Ricci}, {Mushotzky}, {Veilleux}, {Berney}, {Crenshaw},
  {Gehrels}, {Harrison}, {Masetti}, {Soto}, {Stern}, {Treister}, \&
  {Ueda}}]{Oh17}
{Oh}, K., {Schawinski}, K., {Koss}, M., {et~al.} 2017,
  \href{http://dx.doi.org/10.1093/mnras/stw2467}{\JournalTitle{\mnras}, 464,
  1466}

\bibitem[{{Parisi} {et~al.}(2014){Parisi}, {Masetti}, {Rojas},
  {Jim{\'e}nez-Bail{\'o}n}, {Chavushyan}, {Palazzi}, {Bassani}, {Bazzano},
  {Bird}, {Galaz}, {Minniti}, {Morelli}, \& {Ubertini}}]{Parisi14}
{Parisi}, P., {Masetti}, N., {Rojas}, A.~F., {et~al.} 2014,
  \href{http://dx.doi.org/10.1051/0004-6361/201322409}{\JournalTitle{\aap},
  561, A67}

\bibitem[{{Paturel} {et~al.}(2003){Paturel}, {Petit}, {Prugniel}, {Theureau},
  {Rousseau}, {Brouty}, {Dubois}, \& {Cambr{\'e}sy}}]{Paturel03}
{Paturel}, G., {Petit}, C., {Prugniel}, P., {et~al.} 2003,
  \href{http://dx.doi.org/10.1051/0004-6361:20031411}{\JournalTitle{\aap}, 412,
  45}

\bibitem[{{Raimann} {et~al.}(2005){Raimann}, {Storchi-Bergmann}, {Quintana},
  {Hunstead}, \& {Wisotzki}}]{Raimann05}
{Raimann}, D., {Storchi-Bergmann}, T., {Quintana}, H., {Hunstead}, R., \&
  {Wisotzki}, L. 2005,
  \href{http://dx.doi.org/10.1111/j.1365-2966.2005.09665.x}{\JournalTitle{\mnras},
  364, 1239}

\bibitem[{{Ratti} {et~al.}(2010){Ratti}, {Bassa}, {Torres}, {Kuiper},
  {Miller-Jones}, \& {Jonker}}]{Ratti10}
{Ratti}, E.~M., {Bassa}, C.~G., {Torres}, M.~A.~P., {et~al.} 2010,
  \href{http://dx.doi.org/10.1111/j.1365-2966.2010.17252.x}{\JournalTitle{\mnras},
  408, 1866}

\bibitem[{{Ricci} {et~al.}(2015){Ricci}, {Ueda}, {Koss}, {Trakhtenbrot},
  {Bauer}, \& {Gandhi}}]{Ricci15}
{Ricci}, C., {Ueda}, Y., {Koss}, M.~J., {et~al.} 2015,
  \href{http://dx.doi.org/10.1088/2041-8205/815/1/L13}{\JournalTitle{\apjl},
  815, L13}

\bibitem[{{Ricci} {et~al.}(2017{\natexlab{a}}){Ricci}, {Trakhtenbrot}, {Koss},
  {Ueda}, {Delvecchio}, {Treister}, {Schawinski}, {Paltani}, {Oh}, {Lamperti},
  {Berney}, {Gandhi}, {Ichikawa}, {Bauer}, {Ho}, {Asmus}, {Beckmann}, {Soldi},
  {Balokovic}, {Gehrels}, \& {Markwardt}}]{Ricci17a}
{Ricci}, C., {Trakhtenbrot}, B., {Koss}, M.~J., {et~al.} 2017{\natexlab{a}},
  \JournalTitle{ArXiv e-prints},
  \href{http://arxiv.org/abs/1709.03989}{{\sffamily arXiv:1709.03989
  [astro-ph.HE]}}

\bibitem[{{Ricci} {et~al.}(2017{\natexlab{b}}){Ricci}, {Trakhtenbrot}, {Koss},
  {Ueda}, {Schawinski}, {Oh}, {Lamperti}, {Mushotzky}, {Treister}, {Ho},
  {Weigel}, {Bauer}, {Paltani}, {Fabian}, {Xie}, \& {Gehrels}}]{Ricci17b}
---. 2017{\natexlab{b}},
  \href{http://dx.doi.org/10.1038/nature23906}{\JournalTitle{\nat}, 549, 488}

\bibitem[{{Rojas} {et~al.}(2017){Rojas}, {Masetti}, {Minniti},
  {Jim{\'e}nez-Bail{\'o}n}, {Chavushyan}, {Hau}, {McBride}, {Bassani},
  {Bazzano}, {Bird}, {Galaz}, {Gavignaud}, {Landi}, {Malizia}, {Morelli},
  {Palazzi}, {Pati{\~n}o-{\'A}lvarez}, {Stephen}, \& {Ubertini}}]{Rojas17}
{Rojas}, A.~F., {Masetti}, N., {Minniti}, D., {et~al.} 2017,
  \href{http://dx.doi.org/10.1051/0004-6361/201629463}{\JournalTitle{\aap},
  602, A124}

\bibitem[{{Roman} {et~al.}(1996){Roman}, {Nakanishi}, {Tomita}, \&
  {Saito}}]{Roman96}
{Roman}, A.~T., {Nakanishi}, K., {Tomita}, A., \& {Saito}, M. 1996,
  \href{http://dx.doi.org/10.1093/pasj/48.5.679}{\JournalTitle{\pasj}, 48, 679}

\bibitem[{{Roming} {et~al.}(2005){Roming}, {Kennedy}, {Mason}, {Nousek}, {Ahr},
  {Bingham}, {Broos}, {Carter}, {Hancock}, {Huckle}, {Hunsberger}, {Kawakami},
  {Killough}, {Koch}, {McLelland}, {Smith}, {Smith}, {Soto}, {Boyd},
  {Breeveld}, {Holland}, {Ivanushkina}, {Pryzby}, {Still}, \&
  {Stock}}]{Roming05}
{Roming}, P.~W.~A., {Kennedy}, T.~E., {Mason}, K.~O., {et~al.} 2005,
  \href{http://dx.doi.org/10.1007/s11214-005-5095-4}{\JournalTitle{\ssr}, 120,
  95}

\bibitem[{{Rousseau} {et~al.}(2000){Rousseau}, {Paturel}, {Vauglin},
  {Schr{\"o}der}, {de Batz}, {Borsenberger}, {Epchtein}, {Fouqu{\'e}},
  {Kimeswenger}, {Lacombe}, {Le Bertre}, {Mamon}, {Rouan}, {Simon}, \&
  {Tiph{\`e}ne}}]{Rousseau00}
{Rousseau}, J., {Paturel}, G., {Vauglin}, I., {et~al.} 2000,
  \JournalTitle{\aap}, 363, 62

\bibitem[{{Sakamoto} {et~al.}(2008){Sakamoto}, {Barthelmy}, {Barbier},
  {Cummings}, {Fenimore}, {Gehrels}, {Hullinger}, {Krimm}, {Markwardt},
  {Palmer}, {Parsons}, {Sato}, {Stamatikos}, {Tueller}, {Ukwatta}, \&
  {Zhang}}]{Sakamoto08}
{Sakamoto}, T., {Barthelmy}, S.~D., {Barbier}, L., {et~al.} 2008,
  \href{http://dx.doi.org/10.1086/523646}{\JournalTitle{\apjs}, 175, 179}

\bibitem[{{Sakamoto} {et~al.}(2011){Sakamoto}, {Barthelmy}, {Baumgartner},
  {Cummings}, {Fenimore}, {Gehrels}, {Krimm}, {Markwardt}, {Palmer}, {Parsons},
  {Sato}, {Stamatikos}, {Tueller}, {Ukwatta}, \& {Zhang}}]{Sakamoto11}
{Sakamoto}, T., {Barthelmy}, S.~D., {Baumgartner}, W.~H., {et~al.} 2011,
  \href{http://dx.doi.org/10.1088/0067-0049/195/1/2}{\JournalTitle{\apjs}, 195,
  2}

\bibitem[{{Skrutskie} {et~al.}(2006){Skrutskie}, {Cutri}, {Stiening},
  {Weinberg}, {Schneider}, {Carpenter}, {Beichman}, {Capps}, {Chester},
  {Elias}, {Huchra}, {Liebert}, {Lonsdale}, {Monet}, {Price}, {Seitzer},
  {Jarrett}, {Kirkpatrick}, {Gizis}, {Howard}, {Evans}, {Fowler}, {Fullmer},
  {Hurt}, {Light}, {Kopan}, {Marsh}, {McCallon}, {Tam}, {Van Dyk}, \&
  {Wheelock}}]{Skrutskie06}
{Skrutskie}, M.~F., {Cutri}, R.~M., {Stiening}, R., {et~al.} 2006,
  \href{http://dx.doi.org/10.1086/498708}{\JournalTitle{\aj}, 131, 1163}

\bibitem[{{Straal} {et~al.}(2016){Straal}, {Gab{\'a}nyi}, {van Leeuwen},
  {Clarke}, {Dubner}, {Frey}, {Giacani}, \& {Paragi}}]{Straal16}
{Straal}, S.~M., {Gab{\'a}nyi}, K.~{\'E}., {van Leeuwen}, J., {et~al.} 2016,
  \href{http://dx.doi.org/10.3847/0004-637X/822/2/117}{\JournalTitle{\apj},
  822, 117}

\bibitem[{{Sweet} {et~al.}(2014){Sweet}, {Drinkwater}, {Meurer}, {Bekki},
  {Dopita}, {Kilborn}, \& {Nicholls}}]{Sweet14}
{Sweet}, S.~M., {Drinkwater}, M.~J., {Meurer}, G., {et~al.} 2014,
  \href{http://dx.doi.org/10.1088/0004-637X/782/1/35}{\JournalTitle{\apj}, 782,
  35}

\bibitem[{{Tomsick} {et~al.}(2012){Tomsick}, {Bodaghee}, {Chaty}, {Rodriguez},
  {Rahoui}, {Halpern}, {Kalemci}, \& {{\"O}zbey Arabaci}}]{Tomsick12}
{Tomsick}, J.~A., {Bodaghee}, A., {Chaty}, S., {et~al.} 2012,
  \href{http://dx.doi.org/10.1088/0004-637X/754/2/145}{\JournalTitle{\apj},
  754, 145}

\bibitem[{{Tomsick} {et~al.}(2009){Tomsick}, {Chaty}, {Rodriguez}, {Walter}, \&
  {Kaaret}}]{Tomsick09}
{Tomsick}, J.~A., {Chaty}, S., {Rodriguez}, J., {Walter}, R., \& {Kaaret}, P.
  2009,
  \href{http://dx.doi.org/10.1088/0004-637X/701/1/811}{\JournalTitle{\apj},
  701, 811}

\bibitem[{{Tomsick} {et~al.}(2016){Tomsick}, {Krivonos}, {Wang}, {Bodaghee},
  {Chaty}, {Rahoui}, {Rodriguez}, \& {Fornasini}}]{Tomsick16}
{Tomsick}, J.~A., {Krivonos}, R., {Wang}, Q., {et~al.} 2016,
  \href{http://dx.doi.org/10.3847/0004-637X/816/1/38}{\JournalTitle{\apj}, 816,
  38}

\bibitem[{{Trakhtenbrot} {et~al.}(2017){Trakhtenbrot}, {Ricci}, {Koss},
  {Schawinski}, {Mushotzky}, {Ueda}, {Veilleux}, {Lamperti}, {Oh}, {Treister},
  {Stern}, {Harrison}, {Balokovi{\'c}}, \& {Gehrels}}]{Trakhtenbrot17}
{Trakhtenbrot}, B., {Ricci}, C., {Koss}, M.~J., {et~al.} 2017,
  \href{http://dx.doi.org/10.1093/mnras/stx1117}{\JournalTitle{\mnras}, 470,
  800}

\bibitem[{{Truemper}(1982)}]{Truemper82}
{Truemper}, J. 1982,
  \href{http://dx.doi.org/10.1016/0273-1177(82)90070-9}{\JournalTitle{Advances
  in Space Research}, 2, 241}

\bibitem[{{Tueller} {et~al.}(2008){Tueller}, {Mushotzky}, {Barthelmy},
  {Cannizzo}, {Gehrels}, {Markwardt}, {Skinner}, \& {Winter}}]{Tueller08}
{Tueller}, J., {Mushotzky}, R.~F., {Barthelmy}, S., {et~al.} 2008,
  \href{http://dx.doi.org/10.1086/588458}{\JournalTitle{\apj}, 681, 113}

\bibitem[{{Tueller} {et~al.}(2010){Tueller}, {Baumgartner}, {Markwardt},
  {Skinner}, {Mushotzky}, {Ajello}, {Barthelmy}, {Beardmore}, {Brandt},
  {Burrows}, {Chincarini}, {Campana}, {Cummings}, {Cusumano}, {Evans},
  {Fenimore}, {Gehrels}, {Godet}, {Grupe}, {Holland}, {Kennea}, {Krimm},
  {Koss}, {Moretti}, {Mukai}, {Osborne}, {Okajima}, {Pagani}, {Page}, {Palmer},
  {Parsons}, {Schneider}, {Sakamoto}, {Sambruna}, {Sato}, {Stamatikos},
  {Stroh}, {Ukwata}, \& {Winter}}]{Tueller10}
{Tueller}, J., {Baumgartner}, W.~H., {Markwardt}, C.~B., {et~al.} 2010,
  \href{http://dx.doi.org/10.1088/0067-0049/186/2/378}{\JournalTitle{\apjs},
  186, 378}

\bibitem[{{Tzanavaris} {et~al.}(2014){Tzanavaris}, {Gallagher},
  {Hornschemeier}, {Fedotov}, {Eracleous}, {Brandt}, {Desjardins}, {Charlton},
  \& {Gronwall}}]{Tzanavaris14}
{Tzanavaris}, P., {Gallagher}, S.~C., {Hornschemeier}, A.~E., {et~al.} 2014,
  \href{http://dx.doi.org/10.1088/0067-0049/212/1/9}{\JournalTitle{\apjs}, 212,
  9}

\bibitem[{{Ubertini} {et~al.}(2003){Ubertini}, {Lebrun}, {Di Cocco}, {Bazzano},
  {Bird}, {Broenstad}, {Goldwurm}, {La Rosa}, {Labanti}, {Laurent}, {Mirabel},
  {Quadrini}, {Ramsey}, {Reglero}, {Sabau}, {Sacco}, {Staubert}, {Vigroux},
  {Weisskopf}, \& {Zdziarski}}]{Ubertini03}
{Ubertini}, P., {Lebrun}, F., {Di Cocco}, G., {et~al.} 2003,
  \href{http://dx.doi.org/10.1051/0004-6361:20031224}{\JournalTitle{\aap}, 411,
  L131}

\bibitem[{{Uzpen} {et~al.}(2007){Uzpen}, {Kobulnicky}, {Monson}, {Pierce},
  {Clemens}, {Backman}, {Meade}, {Babler}, {Indebetouw}, {Whitney}, {Watson},
  {Wolfire}, {Benjamin}, {Bracker}, {Bania}, {Cohen}, {Cyganowski}, {Devine},
  {Heitsch}, {Jackson}, {Mathis}, {Mercer}, {Povich}, {Rho}, {Robitaille},
  {Sewilo}, {Stolovy}, {Watson}, {Wolff}, \& {Churchwell}}]{Uzpen07}
{Uzpen}, B., {Kobulnicky}, H.~A., {Monson}, A.~J., {et~al.} 2007,
  \href{http://dx.doi.org/10.1086/511736}{\JournalTitle{\apj}, 658, 1264}

\bibitem[{{Vasudevan} {et~al.}(2013){Vasudevan}, {Brandt}, {Mushotzky},
  {Winter}, {Baumgartner}, {Shimizu}, {Schneider}, \& {Nousek}}]{Vasudevan13}
{Vasudevan}, R.~V., {Brandt}, W.~N., {Mushotzky}, R.~F., {et~al.} 2013,
  \href{http://dx.doi.org/10.1088/0004-637X/763/2/111}{\JournalTitle{\apj},
  763, 111}

\bibitem[{{V{\'e}ron-Cetty} \& {V{\'e}ron}(2001)}]{VeronCetty01}
{V{\'e}ron-Cetty}, M.-P., \& {V{\'e}ron}, P. 2001,
  \href{http://dx.doi.org/10.1051/0004-6361:20010718}{\JournalTitle{\aap}, 374,
  92}

\bibitem[{{V{\'e}ron-Cetty} \& {V{\'e}ron}(2010)}]{VeronCetty10}
---. 2010,
  \href{http://dx.doi.org/10.1051/0004-6361/201014188}{\JournalTitle{\aap},
  518, A10}

\bibitem[{{Warwick} {et~al.}(2012){Warwick}, {Saxton}, \& {Read}}]{Warwick12}
{Warwick}, R.~S., {Saxton}, R.~D., \& {Read}, A.~M. 2012,
  \href{http://dx.doi.org/10.1051/0004-6361/201118642}{\JournalTitle{\aap},
  548, A99}

\bibitem[{{Wilson-Hodge} {et~al.}(2011){Wilson-Hodge}, {Cherry}, {Case},
  {Baumgartner}, {Beklen}, {Narayana Bhat}, {Briggs}, {Camero-Arranz},
  {Chaplin}, {Connaughton}, {Finger}, {Gehrels}, {Greiner}, {Jahoda}, {Jenke},
  {Kippen}, {Kouveliotou}, {Krimm}, {Kuulkers}, {Lund}, {Meegan}, {Natalucci},
  {Paciesas}, {Preece}, {Rodi}, {Shaposhnikov}, {Skinner}, {Swartz}, {von
  Kienlin}, {Diehl}, \& {Zhang}}]{Wilson-Hodge11}
{Wilson-Hodge}, C.~A., {Cherry}, M.~L., {Case}, G.~L., {et~al.} 2011,
  \href{http://dx.doi.org/10.1088/2041-8205/727/2/L40}{\JournalTitle{\apjl},
  727, L40}

\bibitem[{{Winkler} {et~al.}(2003){Winkler}, {Courvoisier}, {Di Cocco},
  {Gehrels}, {Gim{\'e}nez}, {Grebenev}, {Hermsen}, {Mas-Hesse}, {Lebrun},
  {Lund}, {Palumbo}, {Paul}, {Roques}, {Schnopper}, {Sch{\"o}nfelder},
  {Sunyaev}, {Teegarden}, {Ubertini}, {Vedrenne}, \& {Dean}}]{Winkler03}
{Winkler}, C., {Courvoisier}, T.~J.-L., {Di Cocco}, G., {et~al.} 2003,
  \href{http://dx.doi.org/10.1051/0004-6361:20031288}{\JournalTitle{\aap}, 411,
  L1}

\bibitem[{{Xiong} {et~al.}(2015){Xiong}, {Zhang}, {Bai}, \& {Zhang}}]{Xiong15}
{Xiong}, D., {Zhang}, X., {Bai}, J., \& {Zhang}, H. 2015,
  \href{http://dx.doi.org/10.1093/mnras/stv812}{\JournalTitle{\mnras}, 450,
  3568}

\end{thebibliography}

\clearpage
\end{document}